\documentclass[reprint,amsmath,amssymb,aps,prb]{revtex4-1}
\usepackage{graphicx}
\usepackage{xcolor}
\usepackage{dcolumn}
\usepackage{bm}
\usepackage{amsmath}
\usepackage[normalem]{ulem}
\usepackage{siunitx}
\usepackage{color,soul}
\usepackage[colorlinks=true, citecolor=red, urlcolor=blue]{hyperref}
\usepackage[para,online,flushleft]{threeparttable}
\makeatletter
\def\blfootnote{\xdef\@thefnmark{}\@footnotetext}
\makeatother

\begin{document}

\preprint{APS/123-QED}
\title{Large Piezoelectric Response of  van der Waals Layered Solids}
\author{Sukriti Manna$^{1,3}$, Prashun Gorai$^{2,3}$, Geoff L. Brennecka$^2$, Cristian V. Ciobanu$^1$, and Vladan Stevanovi\'{c}$^{2,3}$
\footnote{Corresponding author, email: vstevano@mines.edu}}
\affiliation{$^1$Dept. of Mechanical Engineering, Colorado School of Mines, Golden, Colorado 80401, USA\\
$^2$Dept. of Metallurgical and Materials Engineering, Colorado School of Mines, Golden, Colorado 80401, USA\\
$^3$National Renewable Energy Laboratory, Golden, CO 80401, USA}
\date{\today}

\begin{abstract}
The bulk piezoelectric response, as measured by the piezoelectric modulus tensor (\textbf{d}), is determined by a combination of charge redistribution due to strain and the amount of strain produced by the application of stress (stiffness). Motivated by the notion that less stiff materials could exhibit large piezoelectric responses, herein we investigate the piezoelectric modulus of van der Waals-bonded quasi-2D ionic compounds using first-principles calculations. From a pool of 869 known binary and ternary quasi-2D materials, we have identified 135 non-centrosymmetric crystals of which 48 systems are found to have \textbf{d} components larger than the longitudinal piezoelectric modulus of AlN (a common piezoelectric for resonators), and three systems with the response greater than that of PbTiO$_3$, which is among the materials with largest known piezoelectric modulus. None of the identified materials have previously been considered for piezoelectric applications. Furthermore, we find that large \textbf{d} components always couple to the deformations (shearing or axial) of van der Waals ``gaps'' between the layers and are indeed enabled by the weak intra-layer interactions.
\end{abstract}

\keywords{piezoelectricity, quasi-2D, vdW, thin film}

\maketitle

\section{\label{sec:level1}Introduction}

Coupling between the mechanical degrees of freedom and the electric polarization in a solid, a hallmark of piezoelectric materials, has found use in applications that range from sensors, resonators, motors, and actuators, to high-resolution ultrasound devices and miniature filters for cellular communications.\cite{jaffe2012a,uchino1996a,liu2018double} Interestingly, only about 10 
piezoelectric materials, including SiO$_2$ (quartz), LiTaO$_3$, LiNbO$_3$, PZT (lead zirconate titanate)-based, BaTiO$_3$-based, (K,Na)NbO$_3$-based, Bi$_4$Ti$_3$O$_{12}$-based, AlN, and ZnO, are technologically relevant and cover virtually all of these applications.\cite{ieeeweigel, wu2016piezotronics,uchino1998piezoelectric} Expanding the pool of known piezoelectric materials would help broaden the range of applications and allow earth-abundant and non-toxic\cite{saito2004lead} replacements for materials that are presently in use. Furthermore, it would offer more cost- and performance-effective choices beyond a relatively limited set of materials that are at present used for their piezoelectric properties. 

The need for new piezoelectrics was recognized before and was the main motivation behind recent computational efforts in high-throughput screening of inorganic materials for piezoelectric performance. These include the work on perovskite alloys,\cite{armiento_PRB:2014, armiento2011screening} and creation of a database of piezoelectric properties of compounds.\cite{de2015database} Materials with reduced dimensionality, such as 2D mono- and multi-layers have also been investigated recently.\cite{acsnano2017, blonsky2015ab, Li2015,yin2017giant, duerloo2012intrinsic,hinchet2018piezoelectric} In virtually all of these works, the quantity of interest was the piezoelectric coefficient tensor $e_{ij}$, which relates the polarization $P_i$ to strain $\epsilon_{j}$,\cite{nye1985physical,eijdij} (in Voigt notation,\cite{voigt2014lehrbuch} $P_i=\sum_{j}^{}e_{ij}\epsilon_{j}$) and is typically obtained from first-principles calculations. 

\begin{figure}[!t]
\centering
\includegraphics[width=0.85\linewidth]{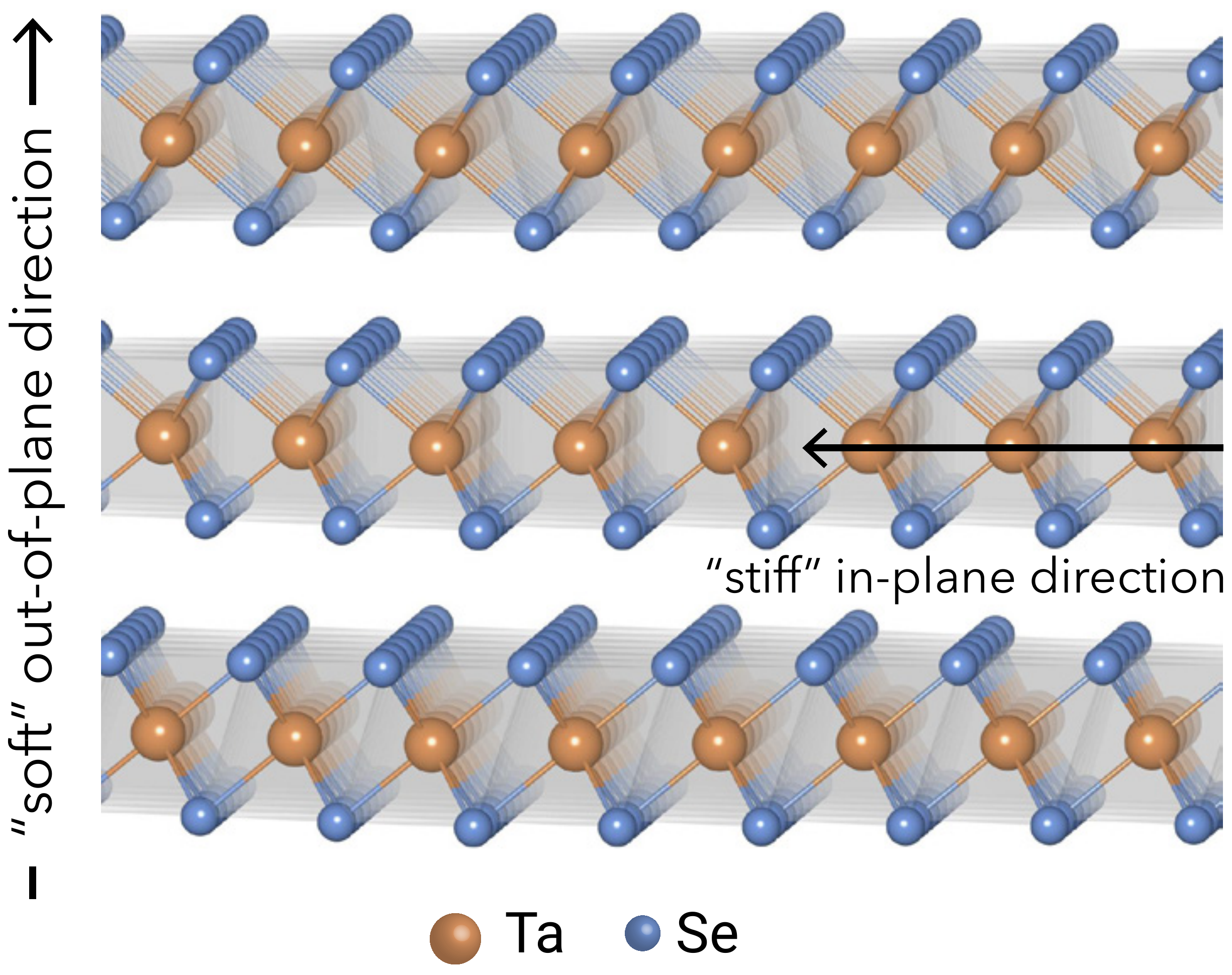}
\caption{Illustration of a quasi-2D material (TaSe$_2$), showing the atomic structure of the layers and the van der Waals spacing between them.}\label{fig:inout}
\end{figure}

In this work, we also apply first-principles calculations to screen for candidate piezoelectrics, but instead of the piezoelectric coefficient tensor we consider the piezoelectric modulus tensor $d_{ij}$, which relates the induced polarization to the applied stress, $P_i=\sum_{j} d_{ij}\sigma_{j}$.\cite{nye1985physical,eijdij} The advantage of using $d_{ij}$ is in that it represents an important figure of merit in a wide range of technological applications\cite{jaffe2012a,uchino1996a} and is a more commonly measured piezoelectric property, especially for materials in the thin film form. Because the induced polarization depends on the applied stress through a combination of the charge redistribution due to strain and the amount of strain that is produced by the applied stress, the piezoelectric modulus tensor $d_{ij}$ depends on both the piezoelectric coefficient tensor $e_{ij}$ and the elastic tensor $C_{ij}$ as:

\begin{equation}
d_{ij} = \sum_{k=1}^{6}e_{ik}(C^{-1})_{kj}.
\label{dce-rel}
\end{equation}

\begin{figure}[!t]
\includegraphics[width=0.95\linewidth]{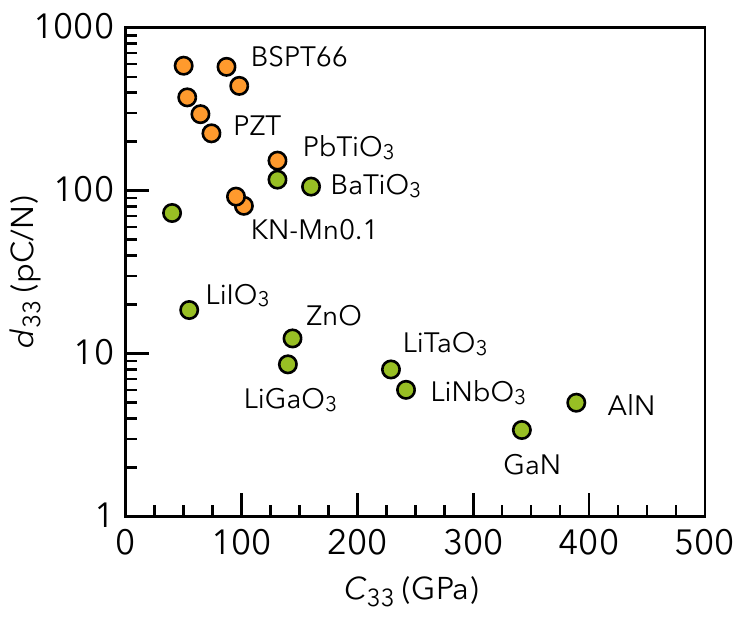}
\caption{Correlation between experimentally measured longitudinal piezoelectric modulus ($d_{33}$) and their elastic modulus ($C_{33}$). Green circles denote data from either single crystals or single-crystal, single-domain samples in case of ferroelectric materials. Orange circles denote polycrystalline samples. Data is sourced from Refs.~\citenum{jaffe2012a, uchino1996a, lec1977elastic, JACE:JACE4240, Lipbtio31993, lefki1994measurement, sotnikov2010elastic, ye2008handbook}.}
\label{fig:d33C33}
\end{figure}
From equation~(\ref{dce-rel}), we can infer that large piezoelectric modulus $d_{ij}$ can be expected in materials with large large $e_{ij}$, but also in systems with low stiffness $C_{ij}$. Softer (less stiff) materials can exhibit large piezoelectric response, as measured by $d_{ij}$, compared to stiffer materials with similar $e_{ij}$. This is illustrated in Figure~\ref{fig:d33C33}, where we notice less stiff materials overall exhibit larger piezoelectric responses in the corresponding direction. 

For this reason, we concentrate our investigation on layered (quasi-2D) van der Waals bonded solids, which can be expected to have relatively low stiffness in the out-of-plane direction due to the presence of weak van der Waals (vdW) interactions between the layers. The questions we are addressing are: (a) whether materials that belong to this class and exhibit strong piezoelectric response ($d_{ij}$) can be found, and (b) if this is true, what is the role of vdW interactions in enabling the strong response. To our knowledge, these questions have not been previously addressed in a systematic way.

Our results confirm the expectations. The search has revealed a number of quasi-2D materials with relatively large $d_{ij}$ components. Out of 869 considered binary and ternary layered VdW systems, we have identified 135 non-centrosymmetric crystals. Out of those we find more than one third (48 compounds) exhibit piezoelectric moduli greater than that of AlN ($d_{33}$=5.5 pC/N), a commonly used piezoelectric material in resonator applications. In addition, we found three layered systems with piezoelectric moduli even larger than that of PbTiO$_3$ ($d_{33}$=119 pC/N ), another established piezoelectric material known for its very large response. It is important to note that none of these vdW systems have been considered previously for piezoelectric applications. 

After performing a thorough analysis of the coupling between various stress and $d_{ij}$ components, we find that in all of these systems, large piezoelectric response is always coupled to the stress components that imply deformations (axial or shear) of the van der Waals ``gaps'' between the layers. This is consistent with the fact that the softest elastic constants are related to these deformations. Ultimately, our results point to the layered vdW systems as a rich chemical space for finding new piezoelectric materials, and introduce elastic properties as additional design criteria for finding materials with large piezoelectric modulus.

\section{Computational Methodology}

In this section, we discuss the details of the computational methodology used in our calculations, broadly divided into three main subsections. The first subsection describes the procedure for identifying quasi-2D structures from the Inorganic Crystal Structure Database (ICSD).\cite{bergerhoff1983inorganic,belsky2002new} Next, we describe the drawbacks of GGA exchange correlation functionals for predicting the properties of layered materials and our approach to overcome this issue. Then, we provide detailed descriptions of our workflow for evaluating the piezoelectric modulus tensors (\textbf{d}) of quasi-2D solids. 

\subsection{Automated Identification of Quasi-2D Materials}\label{q2dsearch}

An essential component of this work is the identification of layered (quasi-2D) vdW materials from ICSD.  To accomplish this, we extend the applications of our previously-developed procedure\cite{gorai2016computational} for automated identification of the binary quasi-2D materials from the ICSD to include ternary chemistries. Similar algorithms have been developed and used for the purpose of broad identification of vdW bonded layered systems by others, including the work of Ashton \textit{et al.},\cite{ashton2017topology} Mounet \textit{et al.},\cite{mounet2016novel} and Cheon \textit{et al.}\cite{cheon2017data} 

Our procedure relies on a slab cutting routine and bond counting. In the first step, we cut out stoichiometric slabs of a certain thickness for all symmetry inequivalent sets of Miller indices $(hkl)$ within a certain range  ($-3\leq h,k,l \leq 3$). Next, for each slab we find the terminations of its surfaces that minimize the number of broken bonds by translating surface atoms from one side of the slab to the other using appropriate lattice vectors. We then count the (minimal) number of broken bonds, {\it i.e.}, the under-coordination, of the surface atoms. The condition of quasi low-dimensional crystals then implies the existence of $(hkl)$ directions for which the corresponding slabs do not have any under-coordinated atom relative to their bulk coordination in the first shell. If there is exactly one such $(hkl)$, the material is a layered material with relatively large spatial gaps separating individual layers. If the number of directions is larger than one, then the corresponding systems are of lower dimensionality, quasi-1D for two such directions and molecular crystals for larger than two. If there are no such directions, the structure is then a connected 3D structure without large spatial gaps.

In Ref.~\citenum{gorai2016computational} we demonstrated the success of our algorithm in searching complex quasi-2D materials, including those with layer stacking in oblique directions, materials with corrugated, accordion-like layers, and those with individual layers composed of multiple atomic layers. Using the described automated algorithm, we have in this work considered $~$3500  binary and $~$8000 additional ternary compounds from the ICSD and classified them into layered (quasi-2D) and not layered materials. We restricted our search to stoichiometric and ordered systems that do not contain rare earth elements and have 50 of less atoms in the unit cell. From our calculations, we have identified 426 binary and 443 ternary quasi-2D compounds. A full list of these materials can be found in supplementary information with their ICSD id.

\subsection{Calculating Piezoelectric Properties of Quasi-2D Materials}\label{VdW}

\begin{figure}[!t]
\includegraphics[width=0.9\linewidth]{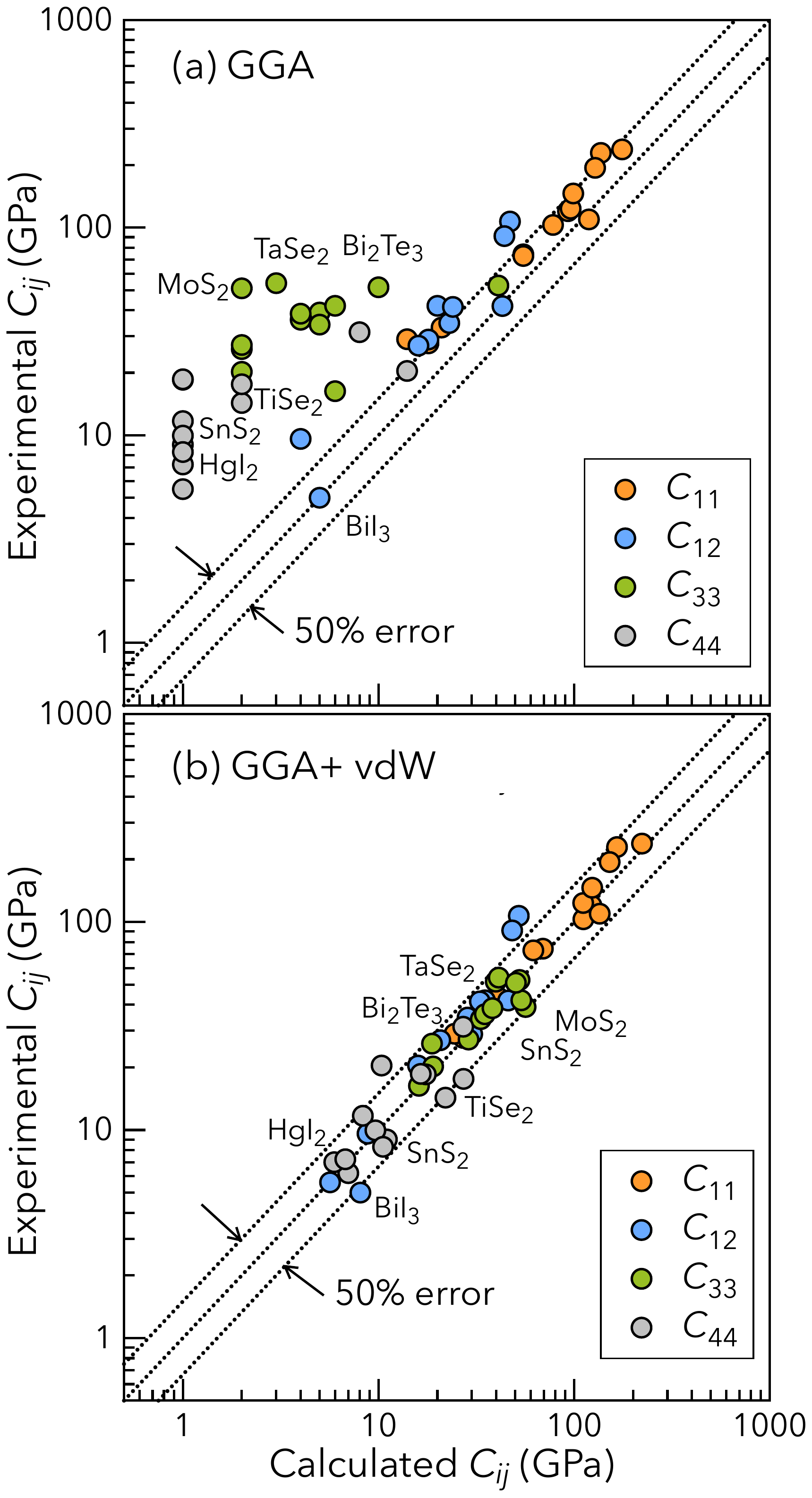}
\caption{Elastic constants of quasi-2D materials calculated with (a) GGA and (b) vdW-corrected functional. The calculated values are compared with experimental ones. The calculations with the vdW-corrected functional are all within 50\% error relative to the measurements. Details of measurement techniques, measurement temperatures, \textit{etc.}, are provided in the supplementary informations.}
\label{elasticcompare}
\end{figure}
\begin{figure*}[!t]
\centering
\includegraphics[width=\linewidth]{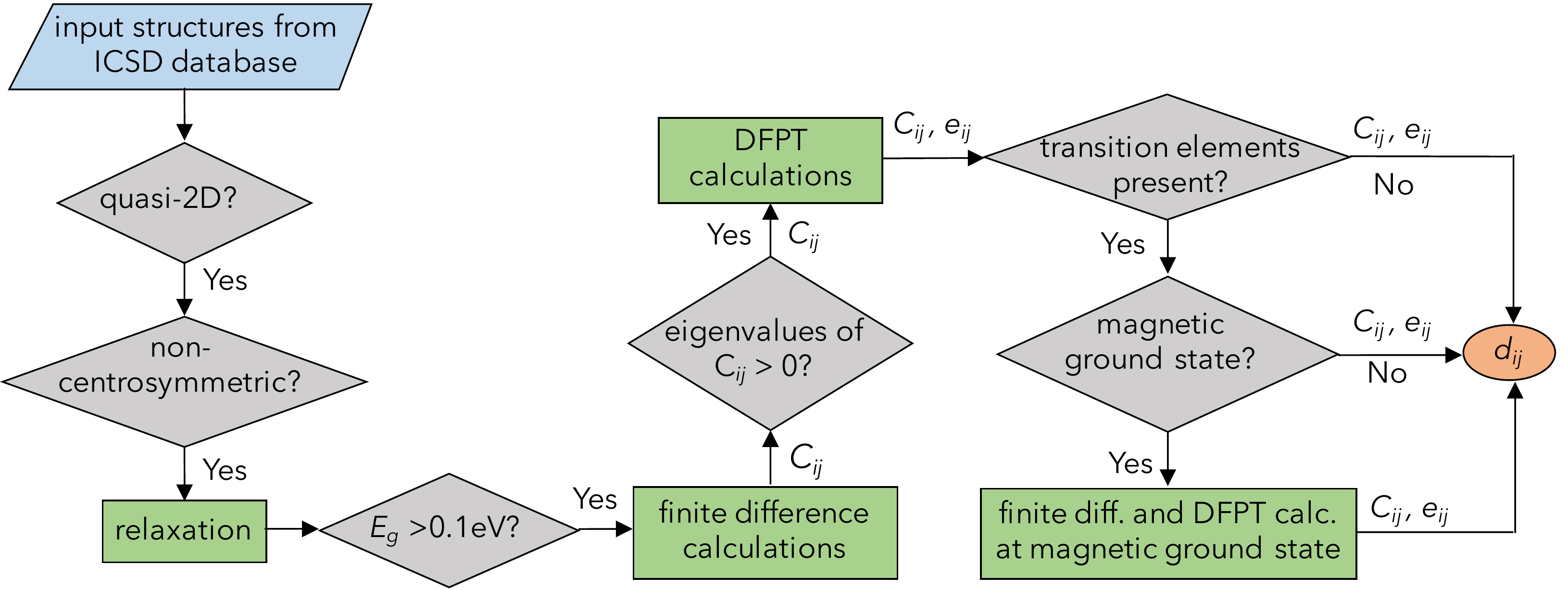}
\caption{Workflow for calculations of piezoelectric coefficient tensor $e_{ij}$, the elastic stiffness $C_{ij}$, and the piezoelectric  modulus tensor $d_{ij}$ of quasi-2D materials. A vdW-corrected functional (optB86) is used in all calculations. We have calculated piezoelectric modulus of 80 quasi-2D materials.}
\label{workflow}
\end{figure*}
%
In quasi-2D structures considered in this work, the individual layers are held together by relatively weak vdW interactions. The standard exchange-correlation functionals typically employed in density functional theory (DFT) calculations, including the calculations of elastic and piezoelectric properties, are known to fail to describe the vdW interactions. This is evident from relatively large errors in the out-of-plane lattice constants and the associated elastic properties.\cite{lebegue2013two} To overcome this issue, we employed a vdW-corrected functional (optB86) as implemented in VASP (Vienna Ab-initio Simulation Package) code\cite{kresse1993ab,kresse1996efficient} to calculate the lattice parameters, elastic, and piezoelectric properties of quasi-2D materials.\cite{PhysRevB.83.195131, klimevs2009chemical} To evaluate the piezoelectric coefficient tensors, we utilize the VASP implementation of the density functional perturbation theory (DFPT)\cite{baroni2001phonons,baroni1987green, gonze1995adiabatic} calculations. A relatively large plane wave cutoff energy of 540 eV is used for structural relaxation, calculation of elastic tensors, and piezoelectric coefficient tensors. A dense k-point grid, defined by $n_{\mathrm{atoms}}$ $\times$ $n_{\mathrm{kpoints}} \approx 1000$, where $n_{\text{atoms}}$ is number of atoms in the primitive cell and $n_{\mathrm{kpoints}}$ is the number of k-points, is employed.  In all our calculations, a very high tolerance of 10$^{-8}$ eV for energy convergence is used, which is an important consideration for conducting DFPT calculations.\cite{de2015database} For calculation of elastic tensors we use a finite difference method. Here, the full elastic tensor is calculated by conducting six finite distortions of the lattice and obtaining elastic constants ($C_{ij}$) from the stress-strain relationship.\cite{le2002symmetry,PRB2010elastic} 

The importance of incorporating vdW-corrections is illustrated in Figure~\ref{elasticcompare}, where we notice significant improvement in predicting elastic constants, particularly $C_{33}$ and $C_{44}$, with vdW-corrected functional (optB86) \cite{PhysRevB.83.195131, klimevs2009chemical} compared to standard GGA-PBE functional.\cite{perdew1996generalized} The GGA-PBE predicted elastic constants are sourced from Ref.\citenum{jain2013commentary}.  A more detailed analysis of the data presented in Figure~\ref{elasticcompare} reveals that the GGA-PBE is still better in predicting in-plane elastic coefficients $C_{11}$ and $C_{12}$, but the error in reproducing $C_{33}$ and $C_{44}$ is a factor of 10 or larger. This is due the fact that these particular two elastic constants are directly related to the deformations of the spatial gaps between the layers and the failure of GGA-PBE in reproducing relatively weak vdW interactions. The comparison of predicted properties with GGA-PBE and vdW-corrected functional is limited only to the elastic constants (Figure~\ref{elasticcompare}); the lack of experimental data on piezoelectric properties of quasi-2D materials prevented us from making similar comparisons for predicted piezoelectric properties. In the supplemental information Table S1, we have provided the experimental details (\textit{e.g.} measurement techniques, measuring temperatures, \textit{etc.}) for each compound shown in Figure~\ref{elasticcompare}. The comparison of calculated and experimental values for the piezoelectric modulus of a few commercially important piezoelectric materials (including AlN and PbTiO$_3$) are shown supplementary information. Predicted piezoelectric moduli are found to be in good agreement with experimental values. 

\subsection{Workflow for identifying quasi-2D piezoelectrics}
A complete workflow we developed for identifying promising quasi-2D piezoelectric materials is illustrated in Figure~\ref{workflow}. The binary and ternary crystal structures from the ICSD database are first screened using the automated algorithm for identifying quasi-2D structures. 
Then, we filter out all centrosymmetric structures based on the space group assigned in ICSD. Out of $\sim$11500 binary and ternary materials we find 869 layered systems, out of which 135 are identified as having non-centrosymmetric structures. Next, the non-centrosymmetric, structures are relaxed using the previously described first-principles calculations employing a vdW-corrected functional. 
As the piezoelectric materials need to have sizable band gaps for their properties to not be screened by the existence of free charge carriers we next employ a band-gap filter. As suggested in the previous works,\cite{gorai2016computational,marom2009describing} for electronic structure calculation, we perform self-consistent GGA-PBE calculations on the vdW-relaxed structures using dense k-point grids. Because of the known band gap error in DFT calculations we use a relatively generous band gap cutoff of 0.1 eV. Fifty additional materials with their band gap smaller than 0.1 eV are discarded as a result. Finite difference calculations are performed  with vdW-corrected functional to obtain the elastic constants (\textbf{C}). Five materials with elastic tensors with negative eigenvalues are also discarded. According to Born stability criteria,\cite{born1940stability} the elastically stable materials always have positive eigenvalues of stiffness matrix. This means that an elastically stable materials always have positive elastic energy for arbitrary homogeneous deformation by an infinitesimal strain.\cite{mouhat2014necessary} With the remaining 80 candidates, we proceed our calculations by performing Density Functional Perturbation Theory (DFPT) \cite{baroni2001phonons,baroni1987green,gonze1995adiabatic} calculations to assess their piezoelectric coefficient tensor, \textbf{e}. Out of the remaining 80 candidates, 38 materials contain transition metal elements. For these, we perform a limited search of the magnetic ground state by enumerating all magnetic configurations in the unit cell and calculating their total energies. The elastic and piezo calculations are then performed only on the lowest energy spin state. This step is necessary because of the strong dependence of the electronic structure and other properties on spin configuration discussed in more details in Ref.~\citenum{gorai2016thermoelectricity}. The automated DFT calculations including initial file generation, calculating properties, data extraction and data handling are performed with the help of PyLada,\cite{pylada} a Python framework for high-throughput first-principles calculations. 

\section{Results and Discussions}

\begin{figure*}[!t]
\centering
\includegraphics[width=\linewidth]{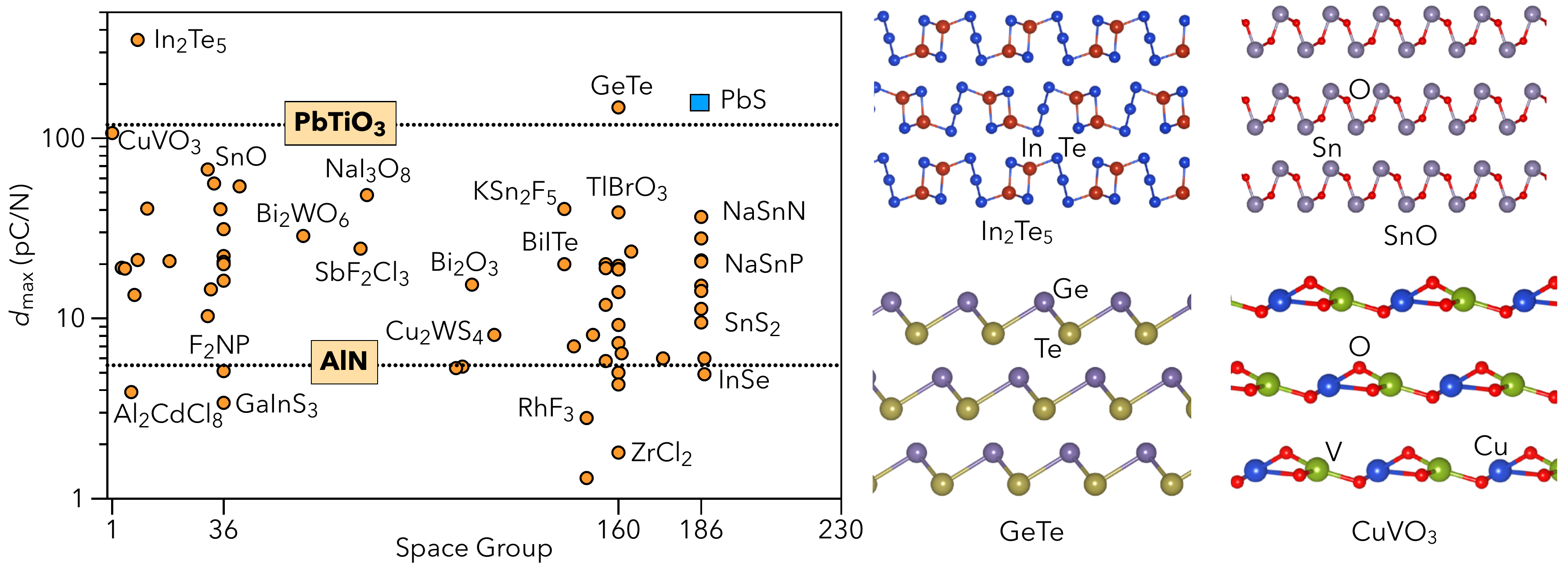}
\caption{Left panel: plot of $d_{max}=max(|d_{ij}|)$ against the space group number for 63 quasi-2D vdW materials with $d_{max}>0$ (out of 135 non-centrosymmetric compounds). The two horizontal dotted lines denote the $d_{33}$ values of PbTiO$_3$ (119 pC/N) and AlN (5.5 pC/N). The blue square in this plot represents the hypothetical structure of PbS (distorted NiAs found in ICSD), not the ground rocksalt structure. Right panel: crystal structures of quasi-2D materials with large piezoelectric moduli.}\label{sgvscandidate}
\end{figure*}

\subsection{Promising Quasi-2D Piezoelectric Materials}
%
Based on the calculated piezoelectric modulus tensor, a number of candidate materials with relatively large $d_{ij}$ components have emerged. They are shown in Figure~\ref{sgvscandidate} with the full list together with the corresponding $d_{ij}$, $C_{ij}$ and $e_{ij}$ values provided in the supplementary information. In addition, the top 20 most promising systems, based on the largest $d_{ij}$ component, are listed in Table~\ref{table1}. 

The piezoelectric modulus tensor (\textbf{d}) is a third rank tensor, and any isotropic averaging scheme will yield zero.\cite{li2000effective} To rank these materials based on the merit of their piezoelectric response, we define $d_{max}=max(|d_{ij}|)$ as the largest element of the absolute ${d}_{ij}$ matrix. 
Then the $d_{max}$ is plotted against the space group number in Figure~\ref{sgvscandidate}. The two reference lines have been drawn for categorizing these candidates -- one representing calculated $d_{33}$ of PbTiO$_3$ and the other representing calculated $d_{33}$ of AlN (see supplementary information for the benchmark agains experiments). 

Please note that PbTiO$_3$ is also a ferroelectric and here we are only using the value for its piezoelectric response. 
More precisely, the PbTiO$_3$ reference line corresponds to the bulk piezoelectric modulus corresponding to the single-crystal single-domain samples.

\begin{table*}
\centering
\begin{threeparttable}[b]
\caption{List of top 20 candidate quasi-2D piezoelectric materials are shown with their space group (SG) number, calculated DFT band gap ($E_g$), maximal piezoelectric modulus $d_{max}$, the $d_{ij}$ component that appears as $d_{max}$, maximal $e_{ij}$ ($e_{max}$), the $e_{max}$ component of $e_{ij}$.}
\setlength{\tabcolsep}{0.8em}
\renewcommand{\arraystretch}{1.25}
\label{candidatelist}
\begin{tabular}{l c c c c c c c c}
\hline \hline
Compound & SG & $E_g$(eV) & $d_{max}$ (pC/N) & max $d_{ij}$ & $e_{max}$ (C/m$^2$) & max $e_{ij}$ & $e_{d_{max}}$ (C/m$^2$)\\
\hline
In$_2$Te$_5$ & 9 & 0.7 & 351.7 & $d_{15}$ & 2.6 & $e_{15}$ & 2.6\\
PbS\tnote{$\dagger$} & 186 & 0.2 & 161.4 & $d_{33}$ & 8.3 & $e_{33}$ & 8.3\\
GeTe  & 160 & 0.6 & 148.4 & $d_{15}$ & 3.3 & $e_{15}$ & 3.3\\
CuVO$_3$ & 1 & 0.9 & 106.9 & $d_{22}$ & 0.8 & $e_{32}$ & 0.2\\
SnO  & 31 & 1.6 & 67.1 & $d_{22}$ & 1.1 & $e_{22}$ & 1.1\\
BiInO$_3$  & 33 & 2.8 & 56.1 & $d_{33}$ & 4.7 & $e_{33}$ & 4.7\\
Bi$_2$WO$_6$\tnote{$\ddagger$}  & 41 & 1.7 & 54.1 & $d_{24}$ & 3.9 & $e_{33}$ & 2.9\\
NaI$_3$O$_8$  & 81 & 2.8 & 48.4 & $d_{14}$ & 0.7 & $e_{31}$ & 0.6\\
NaN$_3$  & 12 & 1.4 & 40.7 & $d_{36}$ & 0.3 & $e_{34}$ & 0.1\\
KSn$_2$F$_5$  & 143 & 3.0 & 40.5 & $d_{15}$ & 0.3 & $e_{15}$ & 0.3\\
MoV$_2$O$_8$  & 35 & 0.8 & 40.4 & $d_{33}$ & 2.9 & $e_{33}$ & 2.9\\
TlBrO$_3$  & 160 & 3.0 & 38.8 & $d_{24}$ & 1.0 & $e_{24}$ & 1.0\\
NaSnN  & 186 & 1.1 & 36.6 & $d_{15}$ & 0.6 & $e_{15}$ & 0.6\\
Cs$_2$Te$_3$\tnote{$\ddagger$}  & 36 & 0.5 &31.3 & $d_{36}$ & 0.6 & $e_{11}$ & 0.4\\
Bi$_2$MoO$_6$  & 61 & 1.7 & 28.7 & $d_{26}$ & 1.6 & $e_{11}$ & 1.3\\
AgI & 186 & 1.3 & 27.8 & $d_{15}$ & 0.3 & $e_{33}$ & 0.1\\
SbF$_2$Cl$_3$  & 79 & 1.5 & 24.4 & $d_{15}$ & 0.2 & $e_{33}$ & 0.1\\
MgCl$_2$  & 115 & 4.9 & 23.5 & $d_{15}$ & 0.2 & $e_{15}$ & 0.2\\
PbRb$_2$O$_3$  & 36 & 1.3 & 22.2 & $d_{34}$ & 0.7 & $e_{34}$ & 0.7\\
BiGeO$_5$  & 9 & 2.3 & 21.1 & $d_{33}$ & 3.0 & $e_{33}$ & 3.0\\

\hline \hline
\end{tabular}\label{table1}
\begin{tablenotes}
     \item[$\dagger$] hypothetical structures
     \item[$\ddagger$] dynamically unstable
   \end{tablenotes}
\end{threeparttable}
\end{table*}

The materials shown in the left panel of Figure~\ref{sgvscandidate} can broadly be divided in three categories. The first category is comprised of quasi-2D compounds with $d_{max}$ larger than the longitudinal piezoelectric modulus of PbTiO$_3$ ($d_{33}$=119 pC/N)\cite{jaffe2012a} -- the key end member of most commercial high-strain piezoelectrics. We found three materials (In$_2$Te$_5$, PbS, and GeTe) in this category. Among them PbS is not in its ground state rocksalt phase, but in the hypothetical distorted NiAs structure which has found its way into the ICSD.\cite{PhysRevB.84.045206} The other two compounds have previously been experimentally synthesized,\cite{chattopadhyay1987neutron,sutherland1976indium} but their piezoelectric moduli have not been reported so far. 

In the second category we group all compounds which have $d_{max}$ larger than the longitudinal piezoelectric modulus of AlN ($d_{33}$=5.5 pC/N)\cite{dubois1999properties} and lower than the longitudinal piezoelectric modulus of PbTiO$_3$. The majority (48 compounds) of the piezoelectric candidates from our study fall in this category revealing that overall the vdW bonded quasi-2D systems indeed exhibit a propensity toward large piezo-response. This group is composed from oxides and other chalcogenides such as CuVO$_3$, SnO, BiInO$_3$, MoV$_2$O$_8$, SnS$_2$, InSe, Cs$_2$Te$_3$ and other (total of 30); halides such as KSn$_2$F$_5$, AgI, MgCl$_2$, and PbI$_2$ (total of 4); pnictides such as NaSnN, NaSnP, KSnAs, and NaN$_3$ (total of 4). Also, a number of materials in this group (10) are the mixed anion systems, \textit{e.g.,} NaI$_3$O$_8$, SbF$_2$Cl$_3$.  Not surprisingly, the more ionic systems like oxides, halides and nitrides are more frequently found closer to the top of the range. Finally, two compounds Cs$_2$Te$_3$ and Bi$_2$WO$_6$ are found in calculations to be dynamically unstable, but both have been experimentally synthesized (likely high-temperature phases).\cite{chuntonov1982synthesis,wolfe1969crystal} 

The last group is composed of materials with the piezoelectric response lower than the longitudinal piezoelectric modulus of AlN. Though these candidates exhibit low piezoelectric response, they could still be useful as the calculated moduli are comparable with that of quartz ($d_{11}$=2.27 pC/N).\cite{bottom1970measurement} We found a total of 12 compounds which fall in this category. Examples include: WS$_2$, RhF$_3$, ZrCl$_2$, and GaInS$_3$. The moduli \textbf{d}, \textbf{e} and \textbf{C} with other informations such as band gap, space group of these compounds are provided in the supplementary information.  

\begin{figure}[!t]
\centering
\includegraphics[width=\linewidth]{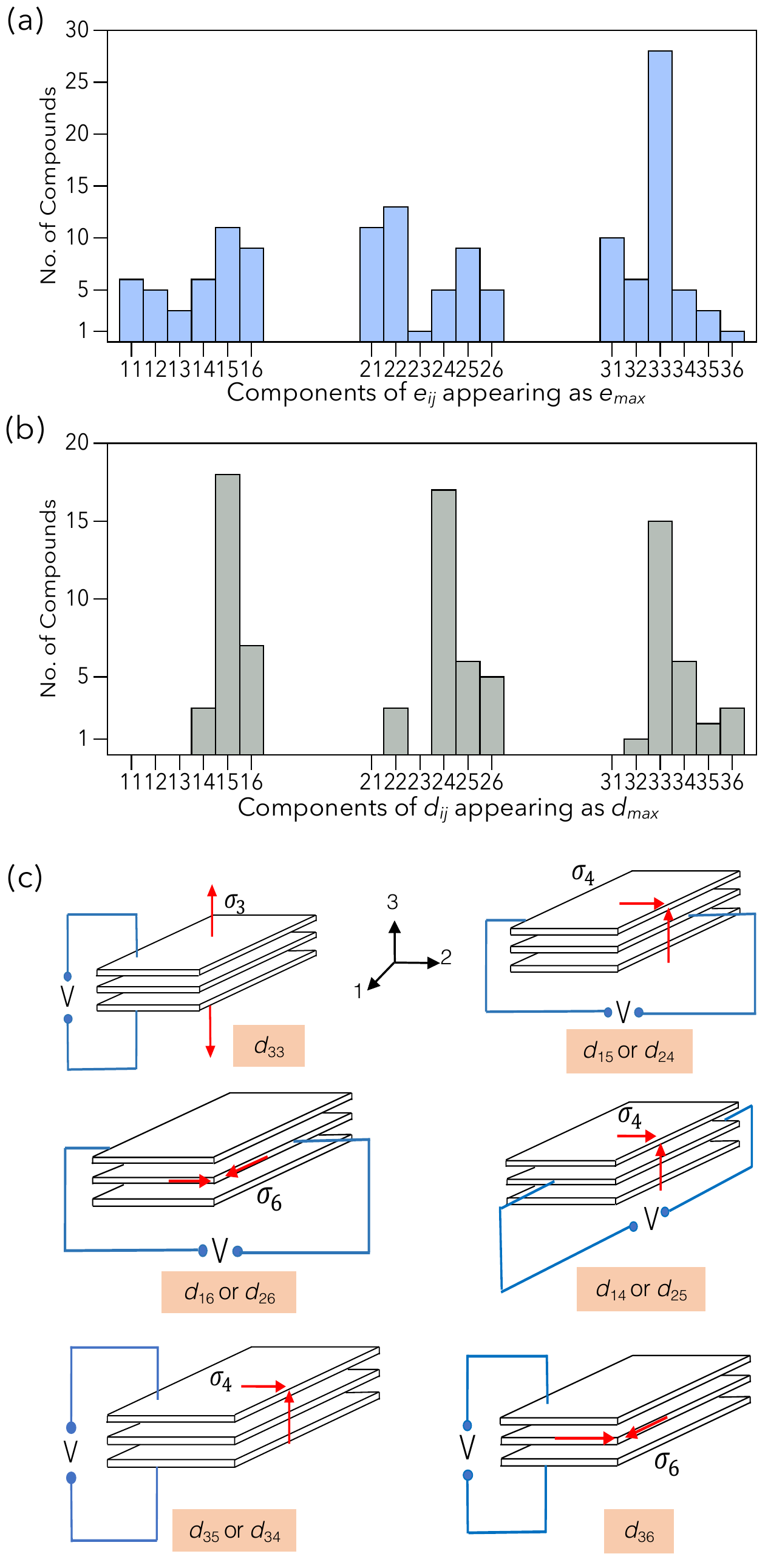}
\caption{Histogram of the maximum components of (a) the piezoelectric coefficient tensor, $e_{max}$  and (b) piezoelectric modulus tensor, $d_{max}$. The most frequent $e_{max}$ is $e_{33}$, whereas for $d_{max}$ the most frequent maximum values are $d_{15}$, $d_{24}$ and $d_{33}$. (c) Schematics of several important piezoelectric operating modes with their corresponding deformation types. The schematics of other relevant piezoelectric operating modes, \textit{i.e.}, components of $d_{ij}$ are shown in the supplementary information.\label{dloc}}
\end{figure}

We also observe that the piezoelectric compounds in the quasi-2D family of solids are clustered mainly in three specific space groups, {\it i.e.}, space group no. 36 (\textit{Cmc2$_1$}), 160 (\textit{R3m}), and 186 (\textit{P6$_3$mc}). This is mainly a reflection of the population bias, as these are the three most frequently occurring non-centrosymmetric space groups in the quasi-2D family of crystals. 

Table~\ref{table1} shows that in 9 of the high-response quasi-2D piezoelectric compounds, the $d_{15}$ component appears as $d_{max}$. Note that because of the freedom in choosing the in-plane axes, $d_{15}$ and $d_{24}$ are virtually indistinguishable (see the discussion section). This component corresponds to the thickness shearing deformation where the material shears like a deck of cards in the in-plane direction, with no change in the other dimension. Materials with large $d_{15}$ can be used in a variety of applications including: sensors, actuators, accelerometer, material testing structural health monitoring, non-destructive testing (NDT), and non-destructive evaluation (NDE).\cite{boivin2017torsional} The components of $d_{ij}$ appearing as $d_{max}$ usually coincide with the components of $e_{ij}$ appearing as $e_{max}$. The distribution of different components of $d_{ij}$ (and $e_{ij}$) appearing as $d_{max}$ (and $e_{max}$) are  discussed in more detail in the next section. 

Another quantity that could influence the piezoelectric response is the band gap of the material. In this work the band gaps are calculated at the DFT level and are also shown in Table \ref{table1} and supplementary information. Only about 1/3 of the studied materials are found to have DFT band gaps below 1 eV. Given the well known underestimation of band gaps in DFT based methods we do not think that materials with DFT band gaps larger than 1 eV would suffer from problems related to the existence of free charge carriers due to thermal fluctuations. However, for those with smaller gaps thermal fluctuation may cause sufficient number of free charge carriers, which may lower the polarization upon straining these materials despite having large piezoelectric moduli. Of course, provided that the real band gap is sufficiently close to the DFT one. For these materials, a more accurate assessment of the electronic structure might be needed before they are considered for applications.

In relation to the chemical composition and toxicity it is also important to not that currently, the most widely used piezoelectric material is lead zirconate titanate (PbZr$_{1-x}$Ti$_x$O$_3$ or PZT).\cite{shrout2007lead} However, PZT causes significant environmental problems because of its high lead content.\cite{shrout2007lead} Hence, significant efforts have been made to develop lead-free piezoelectric materials. \cite{shrout2007lead,takenaka2005current,baettig2005theoretical} In our work, we have identified 44 candidates that do not contain any toxic elements including Pb. Out of 44 candidates, 33 of them have their piezoelectric modulus larger than AlN.

\subsection{Role of van der Waals Interactions}\label{origin}

In order to understand the role of van der Waals interactions on the piezoelectric response of quasi-2D materials, we analyze the relationships of $e_{max}$ and $d_{max}$ to the corresponding strain and stress components, respectively. A histogram showing the number of compounds with a given  $e_{ij}$ component appearing as $e_{max}$ is shown in Figure \ref{dloc} (a). We observe that the most frequent  $e_{max}$ is $e_{33}$. This indicates that in the majority of quasi-2D materials the largest piezoelectric response, as measured by the $e_{ij}$, is along the layer stacking direction. The reason for this behavior is that the relatively weak vdW interactions allow large charge redistribution in the layering direction upon straining the system.   

On the other hand, the piezoelectric modulus tensor \textbf{d} relates relates stress to polarization and combines two types of effects: (1) amount of strain due to application of stress, and (2) amount of charge redistribution (polarization) due to resultant strain produced by the applied stress. A similar histogram of the $d_{ij}$ components appearing as $d_{max}$ is shown in Figure~\ref{dloc}(b). We divide the $d_{ij}$ components into three groups depending on the deformation types (stress components) and the polarization direction. The schematics of the polarization directions and the associated stress components is shown in Figure~\ref{dloc}(c) together with the $d_{ij}$ components connecting the two. Every component of $d_{ij}$ represents a separate piezoelectric operating mode. Schematics of all possible piezoelectric modes are provided in supplementary information.

Group I: Applied stress deforms van der Waals bonds and the measured polarization coincides with the direction of the deformations. The longitudinal mode ($d_{33}$) and shear modes ($d_{15}$ and $d_{24}$) fall in this class. As evident from the histogram in Figure~\ref{dloc}(b) these are the most frequently appearing $d_{max}$ components. In our considered materials, the modes $d_{24}$ and $d_{15}$ are indistinguishable because of the arbitrariness of the choice of axes `1' and `2', while the axis `3' is fixed by layer stacking directions. In both of these modes, the same stress component ($\sigma_4$) is responsible for the deformation, which implies shearing of the van der Waals gaps. On the other hand, in the $d_{33}$, the applied stress axially deforms (stretches or compresses) van der Waals gaps. Hence, the large piezoelectric responses are achieved by deforming (axial or shear) the relatively soft van der Waals bonds. The bar-heights of $d_{15}$ and $d_{24}$ in the histogram of $d_{max}$ in Figure~\ref{dloc}(b) are larger compared to $d_{33}$. This is because the shearing resistance values ($C_{44}$) of quasi-2D materials are lower compared to their axial resistance values ($C_{33}$) (refer to Figure~\ref{elasticcompare}(b)).  

Group II: Applied stress deforms van der Waals bonds, but the measuring polarization directions are different from their deformation directions. The face shear modes ($d_{14}$ and $d_{25}$) and the thickness-extension modes ($d_{31}$ and $d_{32}$) fall in this class. The schematics of $d_{31}$ and $d_{32}$ are provided in the supplementary information. Usually, the direction of polarization is facilitated by the direction of deformation. In such modes, the deformation directions are different from their measured polarization directions. This is the reason behind their low occurrence in the histogram of $d_{max}$ though the van der Waals bond are deformed by the applied stress.

Group III: The van der Waals bond does not deform by the applied stress in these modes. This is why these modes do not appear frequently in the histogram of $d_{max}$. This includes length or width extension modes ($d_{11}$ or $d_{22}$) and shearing modes of type $d_{16}$ and $d_{36}$.

The above discussion implies that the large piezoelectric response is always accompanied and caused by the stress that deforms the ``soft'' van der Waals gaps either through stretching, compression, or shearing. In addition, the analysis of the maximal piezoelectric  moduli and the associated stress components provided guidance to experimentalists of how thin films should be grown to utilize large $d_{max}$. It also describes what kind of mechanical actuation is necessary to achieve large piezoelectric response. This also helps in the design of new devices to take advantage of large $d_{max}$. Finally, these findings can open up a wide variety of devices based on their operation modes or based on new materials for non-conventional modes. For example, materials with large face-shear ($d_{14}$) mode response will be attractive for for torsional applications, like novel gyroscopic sensors or high-precision torsional MEMS actuators.\cite{boivin2017torsional}

\subsection{Axial Piezoelectric and Elastic Anisotropy}\label{aniso}
\begin{figure}[!t]
\includegraphics[width=0.8\linewidth]{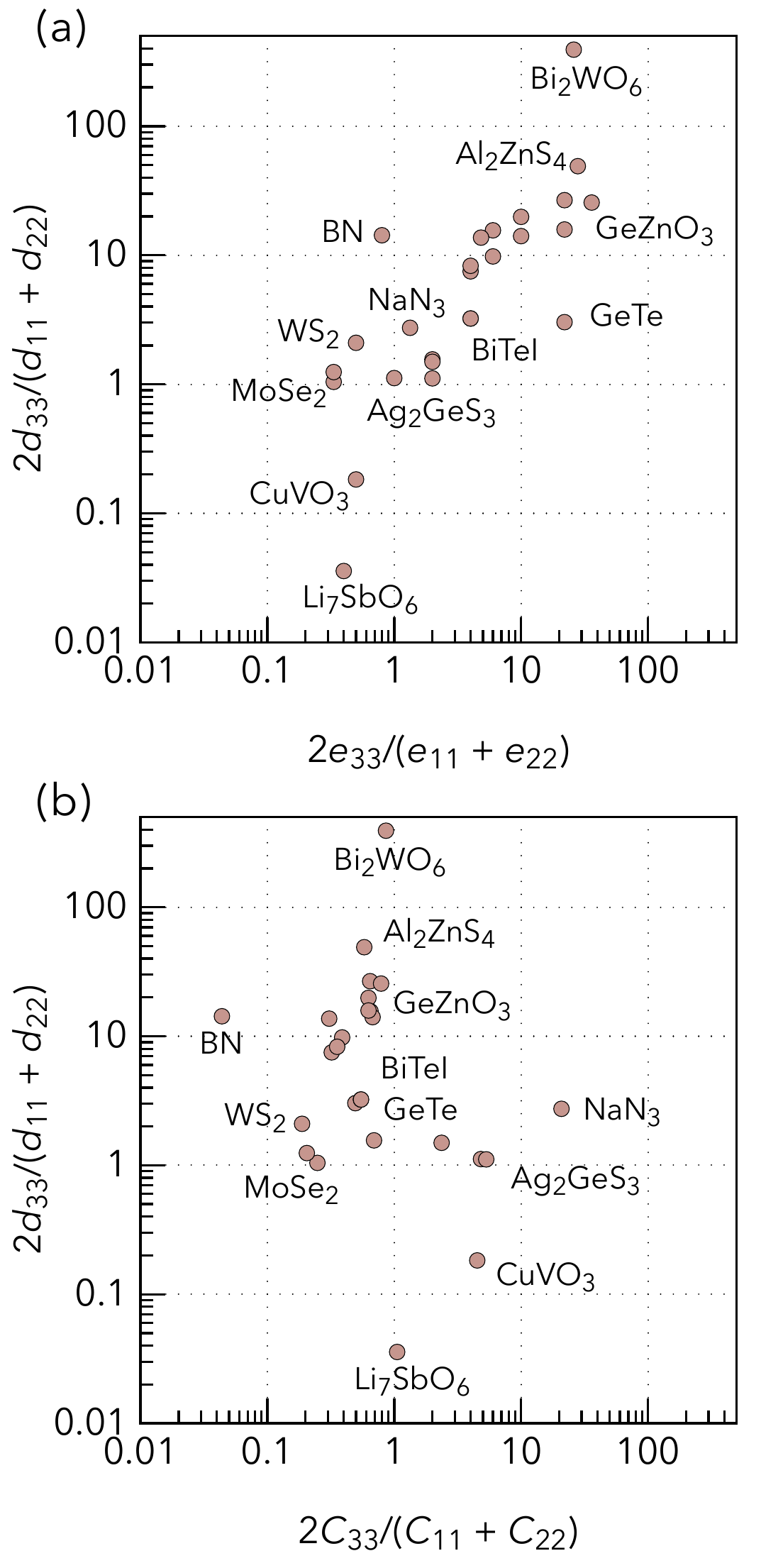}
\caption{Correlation between axial anisotropy, \textit{i.e.}, out-of-plane to in-plane response ratio) of (a) \textbf{d} and \textbf{e} (b) \textbf{d} and \textbf{C}. The results reveal that majority of quasi-2D piezoelectric materials are dominated by out-of-plane piezoelectric responses (both in \textbf{e} and \textbf{d}) but elastically they are dominated in in-plane direction.}\label{Anisotropy-dvse}
\end{figure}
In addition to revealing new candidate piezoelectric materials and explaining the origin of strong piezoelectric response, we have also investigated the anisotropy in axial piezoelectric response in relation to the axial elastic anisotropy. We analyze how the response in the out-of-plane direction compares to the in-plane responses. The in-plane and out-of-plane directions in quasi-2D materials are trivial to define and are illustrated in Figure~\ref{fig:inout}. We define the axial anisotropy in both elastic and piezoelectric responses by the ratio of the out-of-plane component to the in-plane component. Here, the in-plane component is defined by the arithmetic mean of the `11' and `22'-components (invariant to the choice of the in-plane axes), whereas the out-of-plane response is solely defined by the `33'-component. Hence, the axial anisotropy of \textbf{d}, \textbf{e}, and \textbf{C} can be expressed as $2d_{33}/(d_{11} + d_{22})$, $2e_{33}/(e_{11}+e_{22})$, and $2C_{33}/(C_{11}+C_{22})$ respectively. If the axial anisotropy equals or nearly equals 1 then the material is considered to be isotropic in the corresponding quantity responses with respect to the in-plane and out-of-plane directions. If the axial anisotropy is greater (or lower) than 1 the response of a  material to axial deformation is dominated in out-of-plane (in-plane) direction. 

All possible correlations among the axial anisotropy of \textbf{d}, \textbf{e}, and \textbf{C} have been investigated. The correlation between axial anisotropy of \textbf{d} with respect to the axial anisotropy of \textbf{e} and \textbf{C} are shown in Figure~\ref{Anisotropy-dvse} (a) and (b) respectively. From the comparative studies between Figure~\ref{Anisotropy-dvse} (a) and (b), we see that the axial anisotropy of \textbf{d} is mainly dictated by the axial anisotropy of \textbf{e} not by the axial anisotropy of \textbf{C}. The plot between axial anisotropy of \textbf{C} and \textbf{e} is provided in the supplementary information.

From Figure~\ref{Anisotropy-dvse} (a) and (b), we observe that most quasi-2D materials have axial piezoelectric anisotropy parameter (both in \textbf{d} and \textbf{e}) $>$ 1 and axial elastic anisotropy parameter $<$ 1. This implies that the quasi-2D piezoelectric materials are dominant in out-of-plane piezoelectric responses but elastically they are dominant in-plane directions. 
This result corroborates the correlation between elastic softness and large piezoelectric response, \textit{i.e.}, the large piezoelectric responses are observed in elastically softer directions. We found quasi-2D piezoelectric compounds such as CuVO$_3$ and Li$_7$SbO$_6$ are dominated by in-plane piezoelectric response. Compounds such as Bi$_2$WO$_6$, Al$_2$ZnS$_4$, GeZnO$_3$, and GeTe are nearly isotropic in their axial elastic responses but highly anisotropic in axial piezoelectric responses. 

\section{Conclusions}
In conclusion, we performed a large-scale computational (first-principles) assessment of the bulk piezoelectric properties of layered (quasi-2D), vdW bonded materials. In our study we concentrate on the piezoelectric modulus as the measure of the piezoelectric response, which relates mechanical stress and electric polarization and depends on a combination of charge redistribution due to strain and the amount of strain produced by the stress. Overall, out of 135 non-centrosymmetric quasi-2D binary and ternary structures from ICSD we have discovered 51 materials with piezoelectric response larger than that of AlN, a well-known piezoelectric materials used in applications. Out of these 51 systems, we find three with the piezoelectric modulus even larger than that of PbTiO$_3$ that has the piezoelectric modulus among the largest known. More importantly, 33 out of the 51 layered compounds do not contain any toxic elements including Pb. Our results also reveal that the large piezoelectric modulus in vdW systems is directly enabled by the vdW interactions between layers as in majority of compounds the large components of the piezoelectric modulus tensor couple to the stress components that imply deformations (both shear and axial) of the ``soft'' vdW bonds between layers.  
Our results suggest that quasi-2D layered materials are a rich structural space for discovering new piezoelectric materials.

\textit{\textbf{Acknowledgement}}
The authors gratefully acknowledge the support of the National Science Foundation through Grant No. DMREF-1534503. The calculations were performed using the high-performance computing facilities at National Renewable Energy Laboratory (NREL) and at Colorado School of Mines (the Golden Energy Computing Organization) .

\bibliography{biblio}

\begin{thebibliography}{60}%
\makeatletter
\providecommand \@ifxundefined [1]{%
 \@ifx{#1\undefined}
}%
\providecommand \@ifnum [1]{%
 \ifnum #1\expandafter \@firstoftwo
 \else \expandafter \@secondoftwo
 \fi
}%
\providecommand \@ifx [1]{%
 \ifx #1\expandafter \@firstoftwo
 \else \expandafter \@secondoftwo
 \fi
}%
\providecommand \natexlab [1]{#1}%
\providecommand \enquote  [1]{``#1''}%
\providecommand \bibnamefont  [1]{#1}%
\providecommand \bibfnamefont [1]{#1}%
\providecommand \citenamefont [1]{#1}%
\providecommand \href@noop [0]{\@secondoftwo}%
\providecommand \href [0]{\begingroup \@sanitize@url \@href}%
\providecommand \@href[1]{\@@startlink{#1}\@@href}%
\providecommand \@@href[1]{\endgroup#1\@@endlink}%
\providecommand \@sanitize@url [0]{\catcode `\\12\catcode `\$12\catcode
  `\&12\catcode `\#12\catcode `\^12\catcode `\_12\catcode `\%12\relax}%
\providecommand \@@startlink[1]{}%
\providecommand \@@endlink[0]{}%
\providecommand \url  [0]{\begingroup\@sanitize@url \@url }%
\providecommand \@url [1]{\endgroup\@href {#1}{\urlprefix }}%
\providecommand \urlprefix  [0]{URL }%
\providecommand \Eprint [0]{\href }%
\providecommand \doibase [0]{http://dx.doi.org/}%
\providecommand \selectlanguage [0]{\@gobble}%
\providecommand \bibinfo  [0]{\@secondoftwo}%
\providecommand \bibfield  [0]{\@secondoftwo}%
\providecommand \translation [1]{[#1]}%
\providecommand \BibitemOpen [0]{}%
\providecommand \bibitemStop [0]{}%
\providecommand \bibitemNoStop [0]{.\EOS\space}%
\providecommand \EOS [0]{\spacefactor3000\relax}%
\providecommand \BibitemShut  [1]{\csname bibitem#1\endcsname}%
\let\auto@bib@innerbib\@empty
\bibitem [{\citenamefont {Jaffe}(2012)}]{jaffe2012a}%
  \BibitemOpen
  \bibfield  {author} {\bibinfo {author} {\bibfnamefont {B.}~\bibnamefont
  {Jaffe}},\ }\href {https://books.google.com/books?id=GJC\_MBEM4VMC} {\emph
  {\bibinfo {title} {Piezoelectric Ceramics}}},\ Non-Metallic Solids\ (\bibinfo
   {publisher} {Elsevier Science},\ \bibinfo {year} {2012})\BibitemShut
  {NoStop}%
\bibitem [{\citenamefont {Uchino}(1996)}]{uchino1996a}%
  \BibitemOpen
  \bibfield  {author} {\bibinfo {author} {\bibfnamefont {K.}~\bibnamefont
  {Uchino}},\ }\href@noop {} {\emph {\bibinfo {title} {Piezoelectric Actuators
  and Ultrasonic Motors}}},\ Electronic Materials: Science \& Technology\
  (\bibinfo  {publisher} {Springer US},\ \bibinfo {year} {1996})\BibitemShut
  {NoStop}%
\bibitem [{\citenamefont {Liu}\ \emph {et~al.}(2018)\citenamefont {Liu},
  \citenamefont {Wang}, \citenamefont {Wang}, \citenamefont {Cai},
  \citenamefont {Feng}, \citenamefont {Qin},\ and\ \citenamefont
  {Wang}}]{liu2018double}%
  \BibitemOpen
  \bibfield  {author} {\bibinfo {author} {\bibfnamefont {S.}~\bibnamefont
  {Liu}}, \bibinfo {author} {\bibfnamefont {L.}~\bibnamefont {Wang}}, \bibinfo
  {author} {\bibfnamefont {Z.}~\bibnamefont {Wang}}, \bibinfo {author}
  {\bibfnamefont {Y.}~\bibnamefont {Cai}}, \bibinfo {author} {\bibfnamefont
  {X.}~\bibnamefont {Feng}}, \bibinfo {author} {\bibfnamefont {Y.}~\bibnamefont
  {Qin}}, \ and\ \bibinfo {author} {\bibfnamefont {Z.~L.}\ \bibnamefont
  {Wang}},\ }\href {\doibase 10.1021/acsnano.7b08447} {\bibfield  {journal}
  {\bibinfo  {journal} {ACS Nano}\ }\textbf {\bibinfo {volume} {12}},\ \bibinfo
  {pages} {1732} (\bibinfo {year} {2018})},\ \Eprint
  {http://arxiv.org/abs/https://doi.org/10.1021/acsnano.7b08447}
  {https://doi.org/10.1021/acsnano.7b08447} \BibitemShut {NoStop}%
\bibitem [{\citenamefont {Weigel}\ \emph {et~al.}(2002)\citenamefont {Weigel},
  \citenamefont {Morgan}, \citenamefont {Owens}, \citenamefont {Ballato},
  \citenamefont {Lakin}, \citenamefont {Hashimoto},\ and\ \citenamefont
  {Ruppel}}]{ieeeweigel}%
  \BibitemOpen
  \bibfield  {author} {\bibinfo {author} {\bibfnamefont {R.}~\bibnamefont
  {Weigel}}, \bibinfo {author} {\bibfnamefont {D.~P.}\ \bibnamefont {Morgan}},
  \bibinfo {author} {\bibfnamefont {J.~M.}\ \bibnamefont {Owens}}, \bibinfo
  {author} {\bibfnamefont {A.}~\bibnamefont {Ballato}}, \bibinfo {author}
  {\bibfnamefont {K.~M.}\ \bibnamefont {Lakin}}, \bibinfo {author}
  {\bibfnamefont {K.}~\bibnamefont {Hashimoto}}, \ and\ \bibinfo {author}
  {\bibfnamefont {C.~C.~W.}\ \bibnamefont {Ruppel}},\ }\href {\doibase
  10.1109/22.989958} {\bibfield  {journal} {\bibinfo  {journal} {IEEE
  Transactions on Microwave Theory and Techniques}\ }\textbf {\bibinfo {volume}
  {50}},\ \bibinfo {pages} {738} (\bibinfo {year} {2002})}\BibitemShut
  {NoStop}%
\bibitem [{\citenamefont {Wu}\ and\ \citenamefont
  {Wang}(2016)}]{wu2016piezotronics}%
  \BibitemOpen
  \bibfield  {author} {\bibinfo {author} {\bibfnamefont {W.}~\bibnamefont
  {Wu}}\ and\ \bibinfo {author} {\bibfnamefont {Z.~L.}\ \bibnamefont {Wang}},\
  }\href@noop {} {\bibfield  {journal} {\bibinfo  {journal} {Nature Reviews
  Materials}\ }\textbf {\bibinfo {volume} {1}},\ \bibinfo {pages} {16031}
  (\bibinfo {year} {2016})}\BibitemShut {NoStop}%
\bibitem [{\citenamefont {Uchino}(1998)}]{uchino1998piezoelectric}%
  \BibitemOpen
  \bibfield  {author} {\bibinfo {author} {\bibfnamefont {K.}~\bibnamefont
  {Uchino}},\ }\href@noop {} {\bibfield  {journal} {\bibinfo  {journal} {Smart
  Materials and Structures}\ }\textbf {\bibinfo {volume} {7}},\ \bibinfo
  {pages} {273} (\bibinfo {year} {1998})}\BibitemShut {NoStop}%
\bibitem [{\citenamefont {Saito}\ \emph {et~al.}(2004)\citenamefont {Saito},
  \citenamefont {Takao}, \citenamefont {Tani}, \citenamefont {Nonoyama},
  \citenamefont {Takatori}, \citenamefont {Homma}, \citenamefont {Nagaya},\
  and\ \citenamefont {Nakamura}}]{saito2004lead}%
  \BibitemOpen
  \bibfield  {author} {\bibinfo {author} {\bibfnamefont {Y.}~\bibnamefont
  {Saito}}, \bibinfo {author} {\bibfnamefont {H.}~\bibnamefont {Takao}},
  \bibinfo {author} {\bibfnamefont {T.}~\bibnamefont {Tani}}, \bibinfo {author}
  {\bibfnamefont {T.}~\bibnamefont {Nonoyama}}, \bibinfo {author}
  {\bibfnamefont {K.}~\bibnamefont {Takatori}}, \bibinfo {author}
  {\bibfnamefont {T.}~\bibnamefont {Homma}}, \bibinfo {author} {\bibfnamefont
  {T.}~\bibnamefont {Nagaya}}, \ and\ \bibinfo {author} {\bibfnamefont
  {M.}~\bibnamefont {Nakamura}},\ }\href@noop {} {\bibfield  {journal}
  {\bibinfo  {journal} {Nature}\ }\textbf {\bibinfo {volume} {432}},\ \bibinfo
  {pages} {84} (\bibinfo {year} {2004})}\BibitemShut {NoStop}%
\bibitem [{\citenamefont {Armiento}\ \emph {et~al.}(2014)\citenamefont
  {Armiento}, \citenamefont {Kozinsky}, \citenamefont {Hautier}, \citenamefont
  {Fornari},\ and\ \citenamefont {Ceder}}]{armiento_PRB:2014}%
  \BibitemOpen
  \bibfield  {author} {\bibinfo {author} {\bibfnamefont {R.}~\bibnamefont
  {Armiento}}, \bibinfo {author} {\bibfnamefont {B.}~\bibnamefont {Kozinsky}},
  \bibinfo {author} {\bibfnamefont {G.}~\bibnamefont {Hautier}}, \bibinfo
  {author} {\bibfnamefont {M.}~\bibnamefont {Fornari}}, \ and\ \bibinfo
  {author} {\bibfnamefont {G.}~\bibnamefont {Ceder}},\ }\href {\doibase
  10.1103/PhysRevB.89.134103} {\bibfield  {journal} {\bibinfo  {journal}
  {Physical Review B}\ }\textbf {\bibinfo {volume} {89}},\ \bibinfo {pages}
  {134103} (\bibinfo {year} {2014})}\BibitemShut {NoStop}%
\bibitem [{\citenamefont {Armiento}\ \emph {et~al.}(2011)\citenamefont
  {Armiento}, \citenamefont {Kozinsky}, \citenamefont {Fornari},\ and\
  \citenamefont {Ceder}}]{armiento2011screening}%
  \BibitemOpen
  \bibfield  {author} {\bibinfo {author} {\bibfnamefont {R.}~\bibnamefont
  {Armiento}}, \bibinfo {author} {\bibfnamefont {B.}~\bibnamefont {Kozinsky}},
  \bibinfo {author} {\bibfnamefont {M.}~\bibnamefont {Fornari}}, \ and\
  \bibinfo {author} {\bibfnamefont {G.}~\bibnamefont {Ceder}},\ }\href@noop {}
  {\bibfield  {journal} {\bibinfo  {journal} {Physical Review B}\ }\textbf
  {\bibinfo {volume} {84}},\ \bibinfo {pages} {014103} (\bibinfo {year}
  {2011})}\BibitemShut {NoStop}%
\bibitem [{\citenamefont {De~Jong}\ \emph {et~al.}(2015)\citenamefont
  {De~Jong}, \citenamefont {Chen}, \citenamefont {Geerlings}, \citenamefont
  {Asta},\ and\ \citenamefont {Persson}}]{de2015database}%
  \BibitemOpen
  \bibfield  {author} {\bibinfo {author} {\bibfnamefont {M.}~\bibnamefont
  {De~Jong}}, \bibinfo {author} {\bibfnamefont {W.}~\bibnamefont {Chen}},
  \bibinfo {author} {\bibfnamefont {H.}~\bibnamefont {Geerlings}}, \bibinfo
  {author} {\bibfnamefont {M.}~\bibnamefont {Asta}}, \ and\ \bibinfo {author}
  {\bibfnamefont {K.~A.}\ \bibnamefont {Persson}},\ }\href@noop {} {\bibfield
  {journal} {\bibinfo  {journal} {Scientific Data}\ }\textbf {\bibinfo {volume}
  {2}} (\bibinfo {year} {2015})}\BibitemShut {NoStop}%
\bibitem [{\citenamefont {Dong}\ \emph {et~al.}(2017)\citenamefont {Dong},
  \citenamefont {Lou},\ and\ \citenamefont {Shenoy}}]{acsnano2017}%
  \BibitemOpen
  \bibfield  {author} {\bibinfo {author} {\bibfnamefont {L.}~\bibnamefont
  {Dong}}, \bibinfo {author} {\bibfnamefont {J.}~\bibnamefont {Lou}}, \ and\
  \bibinfo {author} {\bibfnamefont {V.~B.}\ \bibnamefont {Shenoy}},\ }\href
  {\doibase 10.1021/acsnano.7b03313} {\bibfield  {journal} {\bibinfo  {journal}
  {ACS Nano}\ }\textbf {\bibinfo {volume} {11}},\ \bibinfo {pages} {8242}
  (\bibinfo {year} {2017})}\BibitemShut {NoStop}%
\bibitem [{\citenamefont {Blonsky}\ \emph {et~al.}(2015)\citenamefont
  {Blonsky}, \citenamefont {Zhuang}, \citenamefont {Singh},\ and\ \citenamefont
  {Hennig}}]{blonsky2015ab}%
  \BibitemOpen
  \bibfield  {author} {\bibinfo {author} {\bibfnamefont {M.~N.}\ \bibnamefont
  {Blonsky}}, \bibinfo {author} {\bibfnamefont {H.~L.}\ \bibnamefont {Zhuang}},
  \bibinfo {author} {\bibfnamefont {A.~K.}\ \bibnamefont {Singh}}, \ and\
  \bibinfo {author} {\bibfnamefont {R.~G.}\ \bibnamefont {Hennig}},\
  }\href@noop {} {\bibfield  {journal} {\bibinfo  {journal} {ACS Nano}\
  }\textbf {\bibinfo {volume} {9}},\ \bibinfo {pages} {9885} (\bibinfo {year}
  {2015})}\BibitemShut {NoStop}%
\bibitem [{\citenamefont {Li}\ and\ \citenamefont {Li}(2015)}]{Li2015}%
  \BibitemOpen
  \bibfield  {author} {\bibinfo {author} {\bibfnamefont {W.}~\bibnamefont
  {Li}}\ and\ \bibinfo {author} {\bibfnamefont {J.}~\bibnamefont {Li}},\ }\href
  {\doibase 10.1007/s12274-015-0878-8} {\bibfield  {journal} {\bibinfo
  {journal} {Nano Research}\ }\textbf {\bibinfo {volume} {8}},\ \bibinfo
  {pages} {3796} (\bibinfo {year} {2015})}\BibitemShut {NoStop}%
\bibitem [{\citenamefont {Yin}\ \emph {et~al.}(2017)\citenamefont {Yin},
  \citenamefont {Gao}, \citenamefont {Zheng}, \citenamefont {Wang},\ and\
  \citenamefont {Ma}}]{yin2017giant}%
  \BibitemOpen
  \bibfield  {author} {\bibinfo {author} {\bibfnamefont {H.}~\bibnamefont
  {Yin}}, \bibinfo {author} {\bibfnamefont {J.}~\bibnamefont {Gao}}, \bibinfo
  {author} {\bibfnamefont {G.~P.}\ \bibnamefont {Zheng}}, \bibinfo {author}
  {\bibfnamefont {Y.}~\bibnamefont {Wang}}, \ and\ \bibinfo {author}
  {\bibfnamefont {Y.}~\bibnamefont {Ma}},\ }\href@noop {} {\bibfield  {journal}
  {\bibinfo  {journal} {The Journal of Physical Chemistry C}\ }\textbf
  {\bibinfo {volume} {121}},\ \bibinfo {pages} {25576} (\bibinfo {year}
  {2017})}\BibitemShut {NoStop}%
\bibitem [{\citenamefont {Duerloo}\ \emph {et~al.}(2012)\citenamefont
  {Duerloo}, \citenamefont {Ong},\ and\ \citenamefont
  {Reed}}]{duerloo2012intrinsic}%
  \BibitemOpen
  \bibfield  {author} {\bibinfo {author} {\bibfnamefont {K.-A.~N.}\
  \bibnamefont {Duerloo}}, \bibinfo {author} {\bibfnamefont {M.~T.}\
  \bibnamefont {Ong}}, \ and\ \bibinfo {author} {\bibfnamefont {E.~J.}\
  \bibnamefont {Reed}},\ }\href@noop {} {\bibfield  {journal} {\bibinfo
  {journal} {The Journal of Physical Chemistry Letters}\ }\textbf {\bibinfo
  {volume} {3}},\ \bibinfo {pages} {2871} (\bibinfo {year} {2012})}\BibitemShut
  {NoStop}%
\bibitem [{\citenamefont {Hinchet}\ \emph {et~al.}(2018)\citenamefont
  {Hinchet}, \citenamefont {Khan}, \citenamefont {Falconi},\ and\ \citenamefont
  {Kim}}]{hinchet2018piezoelectric}%
  \BibitemOpen
  \bibfield  {author} {\bibinfo {author} {\bibfnamefont {R.}~\bibnamefont
  {Hinchet}}, \bibinfo {author} {\bibfnamefont {U.}~\bibnamefont {Khan}},
  \bibinfo {author} {\bibfnamefont {C.}~\bibnamefont {Falconi}}, \ and\
  \bibinfo {author} {\bibfnamefont {S.-W.}\ \bibnamefont {Kim}},\ }\href@noop
  {} {\bibfield  {journal} {\bibinfo  {journal} {Materials Today}\ } (\bibinfo
  {year} {2018})}\BibitemShut {NoStop}%
\bibitem [{\citenamefont {Nye}(1985)}]{nye1985physical}%
  \BibitemOpen
  \bibfield  {author} {\bibinfo {author} {\bibfnamefont {J.~F.}\ \bibnamefont
  {Nye}},\ }\href@noop {} {\emph {\bibinfo {title} {Physical properties of
  crystals: their representation by tensors and matrices}}}\ (\bibinfo
  {publisher} {Oxford University Press},\ \bibinfo {year} {1985})\BibitemShut
  {NoStop}%
\bibitem [{eij()}]{eijdij}%
  \BibitemOpen
  \href@noop {} {\enquote {\bibinfo {title} {We refer \textbf{e}, which couple
  strain and electric displacement as piezoelectric coefficient tensor or
  piezoelectric stress coefficients. \textbf{d}, which couple stress with
  electric displacement, are referred to as piezoelectric modulus tensor or
  piezoelectric strain modulus. in both cases, we express the individual matrix
  elements in standard reduced voigt notation},}\ }\BibitemShut {NoStop}%
\bibitem [{\citenamefont {Voigt}(2014)}]{voigt2014lehrbuch}%
  \BibitemOpen
  \bibfield  {author} {\bibinfo {author} {\bibfnamefont {W.}~\bibnamefont
  {Voigt}},\ }\href@noop {} {\emph {\bibinfo {title} {Lehrbuch der
  kristallphysik (mit ausschluss der kristalloptik)}}}\ (\bibinfo  {publisher}
  {Springer-Verlag},\ \bibinfo {year} {2014})\BibitemShut {NoStop}%
\bibitem [{\citenamefont {Lec}\ and\ \citenamefont
  {Soluch}(1977)}]{lec1977elastic}%
  \BibitemOpen
  \bibfield  {author} {\bibinfo {author} {\bibfnamefont {R.}~\bibnamefont
  {Lec}}\ and\ \bibinfo {author} {\bibfnamefont {W.}~\bibnamefont {Soluch}},\
  }in\ \href@noop {} {\emph {\bibinfo {booktitle} {Ultrasonics Symposium,
  1977}}}\ (\bibinfo {organization} {IEEE},\ \bibinfo {year} {1977})\ pp.\
  \bibinfo {pages} {389--392}\BibitemShut {NoStop}%
\bibitem [{\citenamefont {Pramanick}\ \emph {et~al.}(2011)\citenamefont
  {Pramanick}, \citenamefont {Damjanovic}, \citenamefont {Daniels},
  \citenamefont {Nino},\ and\ \citenamefont {Jones}}]{JACE:JACE4240}%
  \BibitemOpen
  \bibfield  {author} {\bibinfo {author} {\bibfnamefont {A.}~\bibnamefont
  {Pramanick}}, \bibinfo {author} {\bibfnamefont {D.}~\bibnamefont
  {Damjanovic}}, \bibinfo {author} {\bibfnamefont {J.~E.}\ \bibnamefont
  {Daniels}}, \bibinfo {author} {\bibfnamefont {J.~C.}\ \bibnamefont {Nino}}, \
  and\ \bibinfo {author} {\bibfnamefont {J.~L.}\ \bibnamefont {Jones}},\ }\href
  {\doibase 10.1111/j.1551-2916.2010.04240.x} {\bibfield  {journal} {\bibinfo
  {journal} {Journal of the American Ceramic Society}\ }\textbf {\bibinfo
  {volume} {94}},\ \bibinfo {pages} {293} (\bibinfo {year} {2011})}\BibitemShut
  {NoStop}%
\bibitem [{\citenamefont {Li}\ \emph {et~al.}(1993)\citenamefont {Li},
  \citenamefont {Grimsditch}, \citenamefont {Xu},\ and\ \citenamefont
  {Chan}}]{Lipbtio31993}%
  \BibitemOpen
  \bibfield  {author} {\bibinfo {author} {\bibfnamefont {Z.}~\bibnamefont
  {Li}}, \bibinfo {author} {\bibfnamefont {M.}~\bibnamefont {Grimsditch}},
  \bibinfo {author} {\bibfnamefont {X.}~\bibnamefont {Xu}}, \ and\ \bibinfo
  {author} {\bibfnamefont {S.~K.}\ \bibnamefont {Chan}},\ }\href {\doibase
  10.1080/00150199308223459} {\bibfield  {journal} {\bibinfo  {journal}
  {Ferroelectrics}\ }\textbf {\bibinfo {volume} {141}},\ \bibinfo {pages} {313}
  (\bibinfo {year} {1993})}\BibitemShut {NoStop}%
\bibitem [{\citenamefont {Lefki}\ and\ \citenamefont
  {Dormans}(1994)}]{lefki1994measurement}%
  \BibitemOpen
  \bibfield  {author} {\bibinfo {author} {\bibfnamefont {K.}~\bibnamefont
  {Lefki}}\ and\ \bibinfo {author} {\bibfnamefont {G.}~\bibnamefont
  {Dormans}},\ }\href@noop {} {\bibfield  {journal} {\bibinfo  {journal}
  {Journal of Applied Physics}\ }\textbf {\bibinfo {volume} {76}},\ \bibinfo
  {pages} {1764} (\bibinfo {year} {1994})}\BibitemShut {NoStop}%
\bibitem [{\citenamefont {Sotnikov}\ \emph {et~al.}(2010)\citenamefont
  {Sotnikov}, \citenamefont {Schmidt}, \citenamefont {Weihnacht}, \citenamefont
  {Smirnova}, \citenamefont {Chemekova},\ and\ \citenamefont
  {Makarov}}]{sotnikov2010elastic}%
  \BibitemOpen
  \bibfield  {author} {\bibinfo {author} {\bibfnamefont {A.~V.}\ \bibnamefont
  {Sotnikov}}, \bibinfo {author} {\bibfnamefont {H.}~\bibnamefont {Schmidt}},
  \bibinfo {author} {\bibfnamefont {M.}~\bibnamefont {Weihnacht}}, \bibinfo
  {author} {\bibfnamefont {E.~P.}\ \bibnamefont {Smirnova}}, \bibinfo {author}
  {\bibfnamefont {T.~Y.}\ \bibnamefont {Chemekova}}, \ and\ \bibinfo {author}
  {\bibfnamefont {Y.~N.}\ \bibnamefont {Makarov}},\ }\href@noop {} {\bibfield
  {journal} {\bibinfo  {journal} {IEEE transactions on ultrasonics,
  ferroelectrics, and frequency control}\ }\textbf {\bibinfo {volume} {57}}
  (\bibinfo {year} {2010})}\BibitemShut {NoStop}%
\bibitem [{\citenamefont {Ye}(2008)}]{ye2008handbook}%
  \BibitemOpen
  \bibfield  {author} {\bibinfo {author} {\bibfnamefont {Z.-G.}\ \bibnamefont
  {Ye}},\ }\href@noop {} {\emph {\bibinfo {title} {Handbook of advanced
  dielectric, piezoelectric and ferroelectric materials: Synthesis, properties
  and applications}}}\ (\bibinfo  {publisher} {Elsevier},\ \bibinfo {year}
  {2008})\BibitemShut {NoStop}%
\bibitem [{\citenamefont {Bergerhoff}\ \emph {et~al.}(1983)\citenamefont
  {Bergerhoff}, \citenamefont {Hundt}, \citenamefont {Sievers},\ and\
  \citenamefont {Brown}}]{bergerhoff1983inorganic}%
  \BibitemOpen
  \bibfield  {author} {\bibinfo {author} {\bibfnamefont {G.}~\bibnamefont
  {Bergerhoff}}, \bibinfo {author} {\bibfnamefont {R.}~\bibnamefont {Hundt}},
  \bibinfo {author} {\bibfnamefont {R.}~\bibnamefont {Sievers}}, \ and\
  \bibinfo {author} {\bibfnamefont {I.}~\bibnamefont {Brown}},\ }\href@noop {}
  {\bibfield  {journal} {\bibinfo  {journal} {Journal of Chemical Information
  and Computer Sciences}\ }\textbf {\bibinfo {volume} {23}},\ \bibinfo {pages}
  {66} (\bibinfo {year} {1983})}\BibitemShut {NoStop}%
\bibitem [{\citenamefont {Belsky}\ \emph {et~al.}(2002)\citenamefont {Belsky},
  \citenamefont {Hellenbrandt}, \citenamefont {Karen},\ and\ \citenamefont
  {Luksch}}]{belsky2002new}%
  \BibitemOpen
  \bibfield  {author} {\bibinfo {author} {\bibfnamefont {A.}~\bibnamefont
  {Belsky}}, \bibinfo {author} {\bibfnamefont {M.}~\bibnamefont
  {Hellenbrandt}}, \bibinfo {author} {\bibfnamefont {V.~L.}\ \bibnamefont
  {Karen}}, \ and\ \bibinfo {author} {\bibfnamefont {P.}~\bibnamefont
  {Luksch}},\ }\href@noop {} {\bibfield  {journal} {\bibinfo  {journal} {Acta
  Crystallographica Section B: Structural Science}\ }\textbf {\bibinfo {volume}
  {58}},\ \bibinfo {pages} {364} (\bibinfo {year} {2002})}\BibitemShut
  {NoStop}%
\bibitem [{\citenamefont {Gorai}\ \emph
  {et~al.}(2016{\natexlab{a}})\citenamefont {Gorai}, \citenamefont {Toberer},\
  and\ \citenamefont {Stevanovi{\'c}}}]{gorai2016computational}%
  \BibitemOpen
  \bibfield  {author} {\bibinfo {author} {\bibfnamefont {P.}~\bibnamefont
  {Gorai}}, \bibinfo {author} {\bibfnamefont {E.~S.}\ \bibnamefont {Toberer}},
  \ and\ \bibinfo {author} {\bibfnamefont {V.}~\bibnamefont {Stevanovi{\'c}}},\
  }\href@noop {} {\bibfield  {journal} {\bibinfo  {journal} {Journal of
  Materials Chemistry A}\ }\textbf {\bibinfo {volume} {4}},\ \bibinfo {pages}
  {11110} (\bibinfo {year} {2016}{\natexlab{a}})}\BibitemShut {NoStop}%
\bibitem [{\citenamefont {Ashton}\ \emph {et~al.}(2017)\citenamefont {Ashton},
  \citenamefont {Paul}, \citenamefont {Sinnott},\ and\ \citenamefont
  {Hennig}}]{ashton2017topology}%
  \BibitemOpen
  \bibfield  {author} {\bibinfo {author} {\bibfnamefont {M.}~\bibnamefont
  {Ashton}}, \bibinfo {author} {\bibfnamefont {J.}~\bibnamefont {Paul}},
  \bibinfo {author} {\bibfnamefont {S.~B.}\ \bibnamefont {Sinnott}}, \ and\
  \bibinfo {author} {\bibfnamefont {R.~G.}\ \bibnamefont {Hennig}},\ }\href
  {\doibase 10.1103/PhysRevLett.118.106101} {\bibfield  {journal} {\bibinfo
  {journal} {Physical Review Letter}\ }\textbf {\bibinfo {volume} {118}},\
  \bibinfo {pages} {106101} (\bibinfo {year} {2017})}\BibitemShut {NoStop}%
\bibitem [{\citenamefont {Mounet}\ \emph {et~al.}(2018)\citenamefont {Mounet},
  \citenamefont {Gibertini}, \citenamefont {Schwaller}, \citenamefont {Campi},
  \citenamefont {Merkys}, \citenamefont {Marrazzo}, \citenamefont {Sohier},
  \citenamefont {Castelli}, \citenamefont {Cepellotti}, \citenamefont {Pizzi},\
  and\ \citenamefont {Marzari}}]{mounet2016novel}%
  \BibitemOpen
  \bibfield  {author} {\bibinfo {author} {\bibfnamefont {N.}~\bibnamefont
  {Mounet}}, \bibinfo {author} {\bibfnamefont {M.}~\bibnamefont {Gibertini}},
  \bibinfo {author} {\bibfnamefont {P.}~\bibnamefont {Schwaller}}, \bibinfo
  {author} {\bibfnamefont {D.}~\bibnamefont {Campi}}, \bibinfo {author}
  {\bibfnamefont {A.}~\bibnamefont {Merkys}}, \bibinfo {author} {\bibfnamefont
  {A.}~\bibnamefont {Marrazzo}}, \bibinfo {author} {\bibfnamefont
  {T.}~\bibnamefont {Sohier}}, \bibinfo {author} {\bibfnamefont {I.~E.}\
  \bibnamefont {Castelli}}, \bibinfo {author} {\bibfnamefont {A.}~\bibnamefont
  {Cepellotti}}, \bibinfo {author} {\bibfnamefont {G.}~\bibnamefont {Pizzi}}, \
  and\ \bibinfo {author} {\bibfnamefont {N.}~\bibnamefont {Marzari}},\
  }\href@noop {} {\bibfield  {journal} {\bibinfo  {journal} {Nature
  Nanotechnology}\ ,\ \bibinfo {pages} {1}} (\bibinfo {year}
  {2018})}\BibitemShut {NoStop}%
\bibitem [{\citenamefont {Cheon}\ \emph {et~al.}(2017)\citenamefont {Cheon},
  \citenamefont {Duerloo}, \citenamefont {Sendek}, \citenamefont {Porter},
  \citenamefont {Chen},\ and\ \citenamefont {Reed}}]{cheon2017data}%
  \BibitemOpen
  \bibfield  {author} {\bibinfo {author} {\bibfnamefont {G.}~\bibnamefont
  {Cheon}}, \bibinfo {author} {\bibfnamefont {K.-A.~N.}\ \bibnamefont
  {Duerloo}}, \bibinfo {author} {\bibfnamefont {A.~D.}\ \bibnamefont {Sendek}},
  \bibinfo {author} {\bibfnamefont {C.}~\bibnamefont {Porter}}, \bibinfo
  {author} {\bibfnamefont {Y.}~\bibnamefont {Chen}}, \ and\ \bibinfo {author}
  {\bibfnamefont {E.~J.}\ \bibnamefont {Reed}},\ }\href@noop {} {\bibfield
  {journal} {\bibinfo  {journal} {Nano Letters}\ }\textbf {\bibinfo {volume}
  {17}},\ \bibinfo {pages} {1915} (\bibinfo {year} {2017})}\BibitemShut
  {NoStop}%
\bibitem [{\citenamefont {Leb{\`e}gue}\ \emph {et~al.}(2013)\citenamefont
  {Leb{\`e}gue}, \citenamefont {Bj{\"o}rkman}, \citenamefont {Klintenberg},
  \citenamefont {Nieminen},\ and\ \citenamefont {Eriksson}}]{lebegue2013two}%
  \BibitemOpen
  \bibfield  {author} {\bibinfo {author} {\bibfnamefont {S.}~\bibnamefont
  {Leb{\`e}gue}}, \bibinfo {author} {\bibfnamefont {T.}~\bibnamefont
  {Bj{\"o}rkman}}, \bibinfo {author} {\bibfnamefont {M.}~\bibnamefont
  {Klintenberg}}, \bibinfo {author} {\bibfnamefont {R.~M.}\ \bibnamefont
  {Nieminen}}, \ and\ \bibinfo {author} {\bibfnamefont {O.}~\bibnamefont
  {Eriksson}},\ }\href@noop {} {\bibfield  {journal} {\bibinfo  {journal}
  {Physical Review X}\ }\textbf {\bibinfo {volume} {3}},\ \bibinfo {pages}
  {031002} (\bibinfo {year} {2013})}\BibitemShut {NoStop}%
\bibitem [{\citenamefont {Kresse}\ and\ \citenamefont
  {Hafner}(1993)}]{kresse1993ab}%
  \BibitemOpen
  \bibfield  {author} {\bibinfo {author} {\bibfnamefont {G.}~\bibnamefont
  {Kresse}}\ and\ \bibinfo {author} {\bibfnamefont {J.}~\bibnamefont
  {Hafner}},\ }\href@noop {} {\bibfield  {journal} {\bibinfo  {journal}
  {Physical Review B}\ }\textbf {\bibinfo {volume} {47}},\ \bibinfo {pages}
  {558} (\bibinfo {year} {1993})}\BibitemShut {NoStop}%
\bibitem [{\citenamefont {Kresse}\ and\ \citenamefont
  {Furthm{\"u}ller}(1996)}]{kresse1996efficient}%
  \BibitemOpen
  \bibfield  {author} {\bibinfo {author} {\bibfnamefont {G.}~\bibnamefont
  {Kresse}}\ and\ \bibinfo {author} {\bibfnamefont {J.}~\bibnamefont
  {Furthm{\"u}ller}},\ }\href@noop {} {\bibfield  {journal} {\bibinfo
  {journal} {Physical Review B}\ }\textbf {\bibinfo {volume} {54}},\ \bibinfo
  {pages} {11169} (\bibinfo {year} {1996})}\BibitemShut {NoStop}%
\bibitem [{\citenamefont {Klime\ifmmode~\check{s}\else \v{s}\fi{}}\ \emph
  {et~al.}(2011)\citenamefont {Klime\ifmmode~\check{s}\else \v{s}\fi{}},
  \citenamefont {Bowler},\ and\ \citenamefont
  {Michaelides}}]{PhysRevB.83.195131}%
  \BibitemOpen
  \bibfield  {author} {\bibinfo {author} {\bibfnamefont {J.~c.~v.}\
  \bibnamefont {Klime\ifmmode~\check{s}\else \v{s}\fi{}}}, \bibinfo {author}
  {\bibfnamefont {D.~R.}\ \bibnamefont {Bowler}}, \ and\ \bibinfo {author}
  {\bibfnamefont {A.}~\bibnamefont {Michaelides}},\ }\href {\doibase
  10.1103/PhysRevB.83.195131} {\bibfield  {journal} {\bibinfo  {journal}
  {Physical Review B}\ }\textbf {\bibinfo {volume} {83}},\ \bibinfo {pages}
  {195131} (\bibinfo {year} {2011})}\BibitemShut {NoStop}%
\bibitem [{\citenamefont {Klime{\v{s}}}\ \emph {et~al.}(2009)\citenamefont
  {Klime{\v{s}}}, \citenamefont {Bowler},\ and\ \citenamefont
  {Michaelides}}]{klimevs2009chemical}%
  \BibitemOpen
  \bibfield  {author} {\bibinfo {author} {\bibfnamefont {J.}~\bibnamefont
  {Klime{\v{s}}}}, \bibinfo {author} {\bibfnamefont {D.~R.}\ \bibnamefont
  {Bowler}}, \ and\ \bibinfo {author} {\bibfnamefont {A.}~\bibnamefont
  {Michaelides}},\ }\href@noop {} {\bibfield  {journal} {\bibinfo  {journal}
  {Journal of Physics: Condensed Matter}\ }\textbf {\bibinfo {volume} {22}},\
  \bibinfo {pages} {022201} (\bibinfo {year} {2009})}\BibitemShut {NoStop}%
\bibitem [{\citenamefont {Baroni}\ \emph {et~al.}(2001)\citenamefont {Baroni},
  \citenamefont {De~Gironcoli}, \citenamefont {Dal~Corso},\ and\ \citenamefont
  {Giannozzi}}]{baroni2001phonons}%
  \BibitemOpen
  \bibfield  {author} {\bibinfo {author} {\bibfnamefont {S.}~\bibnamefont
  {Baroni}}, \bibinfo {author} {\bibfnamefont {S.}~\bibnamefont
  {De~Gironcoli}}, \bibinfo {author} {\bibfnamefont {A.}~\bibnamefont
  {Dal~Corso}}, \ and\ \bibinfo {author} {\bibfnamefont {P.}~\bibnamefont
  {Giannozzi}},\ }\href@noop {} {\bibfield  {journal} {\bibinfo  {journal}
  {Reviews of Modern Physics}\ }\textbf {\bibinfo {volume} {73}},\ \bibinfo
  {pages} {515} (\bibinfo {year} {2001})}\BibitemShut {NoStop}%
\bibitem [{\citenamefont {Baroni}\ \emph {et~al.}(1987)\citenamefont {Baroni},
  \citenamefont {Giannozzi},\ and\ \citenamefont {Testa}}]{baroni1987green}%
  \BibitemOpen
  \bibfield  {author} {\bibinfo {author} {\bibfnamefont {S.}~\bibnamefont
  {Baroni}}, \bibinfo {author} {\bibfnamefont {P.}~\bibnamefont {Giannozzi}}, \
  and\ \bibinfo {author} {\bibfnamefont {A.}~\bibnamefont {Testa}},\
  }\href@noop {} {\bibfield  {journal} {\bibinfo  {journal} {Physical Review
  Letter}\ }\textbf {\bibinfo {volume} {58}},\ \bibinfo {pages} {1861}
  (\bibinfo {year} {1987})}\BibitemShut {NoStop}%
\bibitem [{\citenamefont {Gonze}(1995)}]{gonze1995adiabatic}%
  \BibitemOpen
  \bibfield  {author} {\bibinfo {author} {\bibfnamefont {X.}~\bibnamefont
  {Gonze}},\ }\href@noop {} {\bibfield  {journal} {\bibinfo  {journal}
  {Physical Review A}\ }\textbf {\bibinfo {volume} {52}},\ \bibinfo {pages}
  {1096} (\bibinfo {year} {1995})}\BibitemShut {NoStop}%
\bibitem [{\citenamefont {Le~Page}\ and\ \citenamefont
  {Saxe}(2002)}]{le2002symmetry}%
  \BibitemOpen
  \bibfield  {author} {\bibinfo {author} {\bibfnamefont {Y.}~\bibnamefont
  {Le~Page}}\ and\ \bibinfo {author} {\bibfnamefont {P.}~\bibnamefont {Saxe}},\
  }\href@noop {} {\bibfield  {journal} {\bibinfo  {journal} {Physical Review
  B}\ }\textbf {\bibinfo {volume} {65}},\ \bibinfo {pages} {104104} (\bibinfo
  {year} {2002})}\BibitemShut {NoStop}%
\bibitem [{\citenamefont {Narayanan}\ \emph {et~al.}(2010)\citenamefont
  {Narayanan}, \citenamefont {Reimanis}, \citenamefont {Fuller~Jr},\ and\
  \citenamefont {Ciobanu}}]{PRB2010elastic}%
  \BibitemOpen
  \bibfield  {author} {\bibinfo {author} {\bibfnamefont {B.}~\bibnamefont
  {Narayanan}}, \bibinfo {author} {\bibfnamefont {I.~E.}\ \bibnamefont
  {Reimanis}}, \bibinfo {author} {\bibfnamefont {E.~R.}\ \bibnamefont
  {Fuller~Jr}}, \ and\ \bibinfo {author} {\bibfnamefont {C.~V.}\ \bibnamefont
  {Ciobanu}},\ }\href@noop {} {\bibfield  {journal} {\bibinfo  {journal}
  {Physical Review B}\ }\textbf {\bibinfo {volume} {81}},\ \bibinfo {pages}
  {104106} (\bibinfo {year} {2010})}\BibitemShut {NoStop}%
\bibitem [{\citenamefont {Perdew}\ \emph {et~al.}(1996)\citenamefont {Perdew},
  \citenamefont {Burke},\ and\ \citenamefont
  {Ernzerhof}}]{perdew1996generalized}%
  \BibitemOpen
  \bibfield  {author} {\bibinfo {author} {\bibfnamefont {J.~P.}\ \bibnamefont
  {Perdew}}, \bibinfo {author} {\bibfnamefont {K.}~\bibnamefont {Burke}}, \
  and\ \bibinfo {author} {\bibfnamefont {M.}~\bibnamefont {Ernzerhof}},\
  }\href@noop {} {\bibfield  {journal} {\bibinfo  {journal} {Physical Review
  Letter}\ }\textbf {\bibinfo {volume} {77}},\ \bibinfo {pages} {3865}
  (\bibinfo {year} {1996})}\BibitemShut {NoStop}%
\bibitem [{\citenamefont {Jain}\ \emph {et~al.}(2013)\citenamefont {Jain},
  \citenamefont {Ong}, \citenamefont {Hautier}, \citenamefont {Chen},
  \citenamefont {Richards}, \citenamefont {Dacek}, \citenamefont {Cholia},
  \citenamefont {Gunter}, \citenamefont {Skinner}, \citenamefont {Ceder},\ and\
  \citenamefont {Persson}}]{jain2013commentary}%
  \BibitemOpen
  \bibfield  {author} {\bibinfo {author} {\bibfnamefont {A.}~\bibnamefont
  {Jain}}, \bibinfo {author} {\bibfnamefont {S.~P.}\ \bibnamefont {Ong}},
  \bibinfo {author} {\bibfnamefont {G.}~\bibnamefont {Hautier}}, \bibinfo
  {author} {\bibfnamefont {W.}~\bibnamefont {Chen}}, \bibinfo {author}
  {\bibfnamefont {W.~D.}\ \bibnamefont {Richards}}, \bibinfo {author}
  {\bibfnamefont {S.}~\bibnamefont {Dacek}}, \bibinfo {author} {\bibfnamefont
  {S.}~\bibnamefont {Cholia}}, \bibinfo {author} {\bibfnamefont
  {D.}~\bibnamefont {Gunter}}, \bibinfo {author} {\bibfnamefont
  {D.}~\bibnamefont {Skinner}}, \bibinfo {author} {\bibfnamefont
  {G.}~\bibnamefont {Ceder}}, \ and\ \bibinfo {author} {\bibfnamefont {K.~A.}\
  \bibnamefont {Persson}},\ }\href {\doibase 10.1063/1.4812323} {\bibfield
  {journal} {\bibinfo  {journal} {APL Materials}\ }\textbf {\bibinfo {volume}
  {1}},\ \bibinfo {pages} {011002} (\bibinfo {year} {2013})},\ \Eprint
  {http://arxiv.org/abs/https://doi.org/10.1063/1.4812323}
  {https://doi.org/10.1063/1.4812323} \BibitemShut {NoStop}%
\bibitem [{\citenamefont {Marom}\ \emph {et~al.}(2009)\citenamefont {Marom},
  \citenamefont {Tkatchenko}, \citenamefont {Scheffler},\ and\ \citenamefont
  {Kronik}}]{marom2009describing}%
  \BibitemOpen
  \bibfield  {author} {\bibinfo {author} {\bibfnamefont {N.}~\bibnamefont
  {Marom}}, \bibinfo {author} {\bibfnamefont {A.}~\bibnamefont {Tkatchenko}},
  \bibinfo {author} {\bibfnamefont {M.}~\bibnamefont {Scheffler}}, \ and\
  \bibinfo {author} {\bibfnamefont {L.}~\bibnamefont {Kronik}},\ }\href@noop {}
  {\bibfield  {journal} {\bibinfo  {journal} {Journal of Chemical Theory and
  Computation}\ }\textbf {\bibinfo {volume} {6}},\ \bibinfo {pages} {81}
  (\bibinfo {year} {2009})}\BibitemShut {NoStop}%
\bibitem [{\citenamefont {Born}(1940)}]{born1940stability}%
  \BibitemOpen
  \bibfield  {author} {\bibinfo {author} {\bibfnamefont {M.}~\bibnamefont
  {Born}},\ }in\ \href@noop {} {\emph {\bibinfo {booktitle} {Mathematical
  Proceedings of the Cambridge Philosophical Society}}},\ Vol.~\bibinfo
  {volume} {36}\ (\bibinfo {organization} {Cambridge University Press},\
  \bibinfo {year} {1940})\ pp.\ \bibinfo {pages} {160--172}\BibitemShut
  {NoStop}%
\bibitem [{\citenamefont {Mouhat}\ and\ \citenamefont
  {Coudert}(2014)}]{mouhat2014necessary}%
  \BibitemOpen
  \bibfield  {author} {\bibinfo {author} {\bibfnamefont {F.}~\bibnamefont
  {Mouhat}}\ and\ \bibinfo {author} {\bibfnamefont {F.-X.}\ \bibnamefont
  {Coudert}},\ }\href@noop {} {\bibfield  {journal} {\bibinfo  {journal}
  {Physical Review B}\ }\textbf {\bibinfo {volume} {90}},\ \bibinfo {pages}
  {224104} (\bibinfo {year} {2014})}\BibitemShut {NoStop}%
\bibitem [{\citenamefont {Gorai}\ \emph
  {et~al.}(2016{\natexlab{b}})\citenamefont {Gorai}, \citenamefont {Toberer},\
  and\ \citenamefont {Stevanovi{\'c}}}]{gorai2016thermoelectricity}%
  \BibitemOpen
  \bibfield  {author} {\bibinfo {author} {\bibfnamefont {P.}~\bibnamefont
  {Gorai}}, \bibinfo {author} {\bibfnamefont {E.~S.}\ \bibnamefont {Toberer}},
  \ and\ \bibinfo {author} {\bibfnamefont {V.}~\bibnamefont {Stevanovi{\'c}}},\
  }\href@noop {} {\bibfield  {journal} {\bibinfo  {journal} {Physical Chemistry
  Chemical Physics}\ }\textbf {\bibinfo {volume} {18}},\ \bibinfo {pages}
  {31777} (\bibinfo {year} {2016}{\natexlab{b}})}\BibitemShut {NoStop}%
\bibitem [{Pylada:()}]{pylada}%
  \BibitemOpen
  Pylada:,\ \href@noop {} {\enquote {\bibinfo {title} {A python framework for
  high-throughput first-principles calculations},}\ }\bibinfo {howpublished}
  {\url{https://github.com/pylada}}\BibitemShut {NoStop}%
\bibitem [{\citenamefont {Li}(2000)}]{li2000effective}%
  \BibitemOpen
  \bibfield  {author} {\bibinfo {author} {\bibfnamefont {J.~Y.}\ \bibnamefont
  {Li}},\ }\href@noop {} {\bibfield  {journal} {\bibinfo  {journal} {Journal of
  the Mechanics and Physics of Solids}\ }\textbf {\bibinfo {volume} {48}},\
  \bibinfo {pages} {529} (\bibinfo {year} {2000})}\BibitemShut {NoStop}%
\bibitem [{\citenamefont {Zagorac}\ \emph {et~al.}(2011)\citenamefont
  {Zagorac}, \citenamefont {Doll}, \citenamefont {Sch\"on},\ and\ \citenamefont
  {Jansen}}]{PhysRevB.84.045206}%
  \BibitemOpen
  \bibfield  {author} {\bibinfo {author} {\bibfnamefont {D.}~\bibnamefont
  {Zagorac}}, \bibinfo {author} {\bibfnamefont {K.}~\bibnamefont {Doll}},
  \bibinfo {author} {\bibfnamefont {J.~C.}\ \bibnamefont {Sch\"on}}, \ and\
  \bibinfo {author} {\bibfnamefont {M.}~\bibnamefont {Jansen}},\ }\href
  {\doibase 10.1103/PhysRevB.84.045206} {\bibfield  {journal} {\bibinfo
  {journal} {Physical Review B}\ }\textbf {\bibinfo {volume} {84}},\ \bibinfo
  {pages} {045206} (\bibinfo {year} {2011})}\BibitemShut {NoStop}%
\bibitem [{\citenamefont {Chattopadhyay}\ \emph {et~al.}(1987)\citenamefont
  {Chattopadhyay}, \citenamefont {Boucherle},\ and\ \citenamefont
  {vonSchnering}}]{chattopadhyay1987neutron}%
  \BibitemOpen
  \bibfield  {author} {\bibinfo {author} {\bibfnamefont {T.}~\bibnamefont
  {Chattopadhyay}}, \bibinfo {author} {\bibfnamefont {J.~X.}\ \bibnamefont
  {Boucherle}}, \ and\ \bibinfo {author} {\bibfnamefont {H.~G.}\ \bibnamefont
  {vonSchnering}},\ }\href {http://stacks.iop.org/0022-3719/20/i=10/a=012}
  {\bibfield  {journal} {\bibinfo  {journal} {Journal of Physics C: Solid State
  Physics}\ }\textbf {\bibinfo {volume} {20}},\ \bibinfo {pages} {1431}
  (\bibinfo {year} {1987})}\BibitemShut {NoStop}%
\bibitem [{\citenamefont {Sutherland}\ \emph {et~al.}(1976)\citenamefont
  {Sutherland}, \citenamefont {Hogg},\ and\ \citenamefont
  {Walton}}]{sutherland1976indium}%
  \BibitemOpen
  \bibfield  {author} {\bibinfo {author} {\bibfnamefont {H.}~\bibnamefont
  {Sutherland}}, \bibinfo {author} {\bibfnamefont {J.}~\bibnamefont {Hogg}}, \
  and\ \bibinfo {author} {\bibfnamefont {P.}~\bibnamefont {Walton}},\
  }\href@noop {} {\bibfield  {journal} {\bibinfo  {journal} {Acta
  Crystallographica Section B: Structural Crystallography and Crystal
  Chemistry}\ }\textbf {\bibinfo {volume} {32}},\ \bibinfo {pages} {2539}
  (\bibinfo {year} {1976})}\BibitemShut {NoStop}%
\bibitem [{\citenamefont {Dubois}\ and\ \citenamefont
  {Muralt}(1999)}]{dubois1999properties}%
  \BibitemOpen
  \bibfield  {author} {\bibinfo {author} {\bibfnamefont {M.-A.}\ \bibnamefont
  {Dubois}}\ and\ \bibinfo {author} {\bibfnamefont {P.}~\bibnamefont
  {Muralt}},\ }\href@noop {} {\bibfield  {journal} {\bibinfo  {journal}
  {Applied Physics Letters}\ }\textbf {\bibinfo {volume} {74}},\ \bibinfo
  {pages} {3032} (\bibinfo {year} {1999})}\BibitemShut {NoStop}%
\bibitem [{\citenamefont {Chuntonov}\ \emph {et~al.}(1982)\citenamefont
  {Chuntonov}, \citenamefont {Orlov}, \citenamefont {Yatsenko}, \citenamefont
  {Grin},\ and\ \citenamefont {Miroshnikova}}]{chuntonov1982synthesis}%
  \BibitemOpen
  \bibfield  {author} {\bibinfo {author} {\bibfnamefont {K.}~\bibnamefont
  {Chuntonov}}, \bibinfo {author} {\bibfnamefont {A.}~\bibnamefont {Orlov}},
  \bibinfo {author} {\bibfnamefont {S.}~\bibnamefont {Yatsenko}}, \bibinfo
  {author} {\bibfnamefont {Y.~N.}\ \bibnamefont {Grin}}, \ and\ \bibinfo
  {author} {\bibfnamefont {L.}~\bibnamefont {Miroshnikova}},\ }\href@noop {}
  {\bibfield  {journal} {\bibinfo  {journal} {Izv. Akad. Nauk SSSR, Neorg.
  Mater}\ }\textbf {\bibinfo {volume} {18}},\ \bibinfo {pages} {1113} (\bibinfo
  {year} {1982})}\BibitemShut {NoStop}%
\bibitem [{\citenamefont {Wolfe}\ \emph {et~al.}(1969)\citenamefont {Wolfe},
  \citenamefont {Newnahm},\ and\ \citenamefont {Kay}}]{wolfe1969crystal}%
  \BibitemOpen
  \bibfield  {author} {\bibinfo {author} {\bibfnamefont {R.}~\bibnamefont
  {Wolfe}}, \bibinfo {author} {\bibfnamefont {R.}~\bibnamefont {Newnahm}}, \
  and\ \bibinfo {author} {\bibfnamefont {M.}~\bibnamefont {Kay}},\ }\href@noop
  {} {\bibfield  {journal} {\bibinfo  {journal} {Solid State Communications}\
  }\textbf {\bibinfo {volume} {7}},\ \bibinfo {pages} {1797} (\bibinfo {year}
  {1969})}\BibitemShut {NoStop}%
\bibitem [{\citenamefont {Bottom}(1970)}]{bottom1970measurement}%
  \BibitemOpen
  \bibfield  {author} {\bibinfo {author} {\bibfnamefont {V.~E.}\ \bibnamefont
  {Bottom}},\ }\href@noop {} {\bibfield  {journal} {\bibinfo  {journal}
  {Journal of Applied physics}\ }\textbf {\bibinfo {volume} {41}},\ \bibinfo
  {pages} {3941} (\bibinfo {year} {1970})}\BibitemShut {NoStop}%
\bibitem [{\citenamefont {Boivin}\ \emph {et~al.}(2017)\citenamefont {Boivin},
  \citenamefont {B{\'{e}}langer},\ and\ \citenamefont
  {Zednik}}]{boivin2017torsional}%
  \BibitemOpen
  \bibfield  {author} {\bibinfo {author} {\bibfnamefont {G.}~\bibnamefont
  {Boivin}}, \bibinfo {author} {\bibfnamefont {P.}~\bibnamefont
  {B{\'{e}}langer}}, \ and\ \bibinfo {author} {\bibfnamefont {R.~J.}\
  \bibnamefont {Zednik}},\ }in\ \href {\doibase
  10.4028/www.scientific.net/MSF.879.637} {\emph {\bibinfo {booktitle} {THERMEC
  2016}}},\ \bibinfo {series} {Materials Science Forum}, Vol.\ \bibinfo
  {volume} {879}\ (\bibinfo  {publisher} {Trans Tech Publications},\ \bibinfo
  {year} {2017})\ pp.\ \bibinfo {pages} {637--641}\BibitemShut {NoStop}%
\bibitem [{\citenamefont {Shrout}\ and\ \citenamefont
  {Zhang}(2007)}]{shrout2007lead}%
  \BibitemOpen
  \bibfield  {author} {\bibinfo {author} {\bibfnamefont {T.~R.}\ \bibnamefont
  {Shrout}}\ and\ \bibinfo {author} {\bibfnamefont {S.~J.}\ \bibnamefont
  {Zhang}},\ }\href@noop {} {\bibfield  {journal} {\bibinfo  {journal} {Journal
  of Electroceramics}\ }\textbf {\bibinfo {volume} {19}},\ \bibinfo {pages}
  {113} (\bibinfo {year} {2007})}\BibitemShut {NoStop}%
\bibitem [{\citenamefont {Takenaka}\ and\ \citenamefont
  {Nagata}(2005)}]{takenaka2005current}%
  \BibitemOpen
  \bibfield  {author} {\bibinfo {author} {\bibfnamefont {T.}~\bibnamefont
  {Takenaka}}\ and\ \bibinfo {author} {\bibfnamefont {H.}~\bibnamefont
  {Nagata}},\ }\href@noop {} {\bibfield  {journal} {\bibinfo  {journal}
  {Journal of the European Ceramic Society}\ }\textbf {\bibinfo {volume}
  {25}},\ \bibinfo {pages} {2693} (\bibinfo {year} {2005})}\BibitemShut
  {NoStop}%
\bibitem [{\citenamefont {Baettig}\ \emph {et~al.}(2005)\citenamefont
  {Baettig}, \citenamefont {Schelle}, \citenamefont {LeSar}, \citenamefont
  {Waghmare},\ and\ \citenamefont {Spaldin}}]{baettig2005theoretical}%
  \BibitemOpen
  \bibfield  {author} {\bibinfo {author} {\bibfnamefont {P.}~\bibnamefont
  {Baettig}}, \bibinfo {author} {\bibfnamefont {C.~F.}\ \bibnamefont
  {Schelle}}, \bibinfo {author} {\bibfnamefont {R.}~\bibnamefont {LeSar}},
  \bibinfo {author} {\bibfnamefont {U.~V.}\ \bibnamefont {Waghmare}}, \ and\
  \bibinfo {author} {\bibfnamefont {N.~A.}\ \bibnamefont {Spaldin}},\
  }\href@noop {} {\bibfield  {journal} {\bibinfo  {journal} {Chemistry of
  materials}\ }\textbf {\bibinfo {volume} {17}},\ \bibinfo {pages} {1376}
  (\bibinfo {year} {2005})}\BibitemShut {NoStop}%
\end{thebibliography}%
\end{document}